\documentclass[superscriptaddress,twocolumn,english,10pt,prb,showpacs]{revtex4-1}
\usepackage[latin9]{inputenc}
\usepackage[T1]{fontenc}
\usepackage{amsmath}
\usepackage{amssymb}
\usepackage{graphicx}
\usepackage{babel}
\usepackage{float}
\usepackage[colorlinks]{hyperref}
\usepackage{color}

\providecommand{\tabularnewline}{\\}

\global\long\def\ket#1{\left| #1\right\rangle }
\global\long\def\bra#1{\left\langle #1 \right|}

\global\long\def\av#1{\left\langle #1 \right\rangle }
\global\long\def\tr{\text{tr}}
\global\long\def\Tr{\text{Tr}}
\global\long\def\pd{\partial}
\global\long\def\im{\text{Im}}
\global\long\def\re{\text{Re}}
\global\long\def\sgn{\text{sgn}}

\global\long\def\abs#1{\left|#1\right|}

\global\long\def\bs#1{\boldsymbol{#1}}
\global\long\def\t#1{\text{#1}}
\definecolor{applegreen}{rgb}{0.55, 0.71, 0.0}
\definecolor{TRRR}{rgb}{1, 0, 0.0}
\definecolor{prorange}{rgb}{1., .5, 0.0}

\begin{document}
\title{Nonequilibrium phases and phase transitions of the XY-model}

\author{Tharnier O. Puel}
\email{tharnier@me.com}
\address{Zhejiang Institute of Modern Physics and Department of Physics, Zhejiang University, Hangzhou,
Zhejiang 310027, China}
\affiliation{Zhejiang Province Key Laboratory of Quantum Technology and Device, Zhejiang University, Hangzhou 310027, China}

\author{Stefano Chesi}
\email{stefano.chesi@csrc.ac.cn}
\address{Beijing Computational Science Research Center, Beijing 100193, China}
\affiliation{Department of Physics, Beijing Normal University, Beijing 100875, China}

\author{Stefan Kirchner}
\email{stefan.kirchner@correlated-matter.com}
\address{Zhejiang Institute of Modern Physics and Department of Physics, Zhejiang University, Hangzhou,
Zhejiang 310027, China}
\affiliation{Zhejiang Province Key Laboratory of Quantum Technology and Device, Zhejiang University, Hangzhou 310027, China}

\author{Pedro Ribeiro}
\email{pedrojgribeiro@tecnico.ulisboa.pt}
\address{CeFEMA, Instituto Superior T\'{e}cnico, Universidade de Lisboa Av. Rovisco
Pais, 1049-001 Lisboa, Portugal}
\affiliation{Beijing Computational Science Research Center, Beijing 100193, China}

\begin{abstract}
We obtain the steady-state phase diagram of a
transverse field XY spin chain coupled at its ends to magnetic reservoirs held at different magnetic potentials. In the long-time limit, the magnetization bias across the system generates a current-carrying non-equilibrium steady-state. We characterize the different non-equilibrium phases as functions of the chain's parameters and magnetic potentials, in terms of their correlation functions and entanglement content.  
The mixed-order transition, recently observed for the particular case of a transverse field Ising chain, is established to emerge as a generic out-of-equilibrium feature and its critical exponents are determined analytically. 
Results are also contrasted with those obtained in the limit of Markovian reservoirs. Our findings should prove helpful in establishing the properties of non-equilibrium phases and phase transitions of extended open quantum systems. 
\end{abstract}

\maketitle


\section{Introduction}

Quantum matter out of thermal equilibrium has become a central research topic in recent years. An important class of problems deal with non-equilibrium quantum states of systems that are in contact with multiple baths which in turn are held at  specified thermodynamic potentials.
Such states are not bounded by equilibrium fluctuation relations and thus may host phases of matter that are impossible to realize in equilibrium. Therefore, phase changes far from equilibrium may exist that lack equilibrium counterparts.

Far-from-equilibrium quantum states are routinely realized in mesoscopic solid-state devices \cite{Pothier1997,Anthore.03,Chen.09} and recently have also become available in cold atomic gas settings \cite{Brantut2013}. It thus is timely to explore  the properties of phases of current-carrying matter and address the conditions which have to be met for their emergence. 

Non-equilibrium transport across quantum materials dates back to Landauer and B\"{u}ttiker \cite{PhysRevLett.49.1739}, who were motivated by the failure of semi-classical Boltzmann-like approaches to understand phenomena such as the conductance quantization across mesoscopic conductors. 
For non-interacting systems,  quantum transport is by now well understood \cite{stefanucci_vanleeuwen_2013,datta1997electronic,imry2002introduction}.  However, in systems where the physical properties are determined by the electron-electron interaction,  progress has been much slower. 
Here, one often has to resort to either approximate methods or  numerically exact techniques \cite{RevModPhys.86.779} which, however,  are often restricted to small systems or comparatively high temperatures. Exact analytical results, available for integrable models in one dimension, do not typically generalize to open setups. Moreover, non-thermal steady-states in Luttinger liquids \cite{Gutman.2008,Gutman.2009,Dinh.2010}, seem to be less general than their equilibrium counterparts. 

Considerable progress has been made in the  Markovian case, where the environment lacks memory \cite{Prosen2008,Prosen2010a,Prosen2011a,Prosen2014}.  The applicability of the Markovian case is, however, limited to  extreme non-equilibrium conditions (e.g., very large bias or temperature) and is of restricted  use for realistic transport setups \cite{Ribeiro2014e,Ribeiro2015f}.  

Other  recent developments to study transport include, the study of so-called generalized hydrodynamic methods available for integrable systems \cite{PhysRevLett.117.207201,PhysRevX.6.041065} and hybrid approaches involving Lindblad dynamics \cite{Arrigoni2013}. However, these methods are not yet able to describe current-carrying steady-states in extended mesoscopic systems.

Our recent analysis of the exactly solvable transverse field Ising chain attached to  macroscopic reservoirs
has allowed us study a symmetry-breaking quantum phase transition in the  steady-state of an extended non-equilibrium system \cite{Puel-Chesi-Kirchner-Ribeiro-2019}.
At the equilibrium level, this model can be mapped onto that of  non-interacting fermions via  a Jordan-Wigner transformation and is thus solvable by elementary means.

The non-thermal steady-state of this model is, however,   much richer and allows for a peculiar symmetry-breaking quantum phase transition.
In particular, we have shown that this transition to be of mixed-order (or hybrid) nature, with a discontinuous order parameter and diverging correlation length. This type of transitions were first discussed by  Thouless in 1969 \citep{Thouless.69} in the context of classical spin chains with long-range interactions and have since then reported in different environments \citep{PhysRevLett.112.015701,PhysRevE.93.012124,PhysRevE.94.062126,PhysRevE.95.022109,Alert12906}.

Even though realistic systems are only approximately described by exactly solvable models at best, exact solutions are still of considerable value. Not only can they be important in unveiling features of 
novel effects but they are commonly instrumental in  
benchmarking numerical and approximate methods.
Therefore, exact solutions are particularly helpful in situations where no reliable numerical or approximate methods yet exist, such as in the description of current-carrying steady-states of interacting systems.

In this article, we provide a set of exact results of  steady-state phases and phase transitions of an  $XY$ spin chain in a transverse field  coupled to magnetic reservoirs held at different magnetizations. Our analysis extends and  generalizes the findings in Ref. [\onlinecite{Puel-Chesi-Kirchner-Ribeiro-2019}] and points out new regimes that are not present in the Ising case. In the Markovian limit, we recover previous results obtained for $XY$ spin chains coupled to free-of-memory reservoirs \citep{Prosen2008,_unkovi__2010,Ajisaka_2014} where an out-of-equilibrium phase transition with spontaneous emergence of long-range order  has been found.

The paper is organized as follows. 
In section \ref{sec:Model-and-Method} we define the out-of-equilibrium model and briefly describe the methods used to solve it. 
In section \ref{sec:Phase-diagram} we describe in detail the non-equilibrium phase diagram based on the energy current and the properties of the occupation number.
The correlation functions in the various phases are analysed in section \ref{sec: mixed-order phase transition}, where we also discuss the critical behavior at the mixed-order phase transition and the characteristic oscillations in the $z$-correlation function.
Universal features of the entanglement entropy are discussed in section \ref{sec:Entropy}. 
Finally, we summarize and conclude our work in section \ref{sec:Discussions}.

\section{Model and Method\label{sec:Model-and-Method}}

\subsection{Hamiltonian and Jordan-Wigner mapping}

We consider an $XY$-spin chain of $N$ sites (labeled by $r$), exchange coupling $J$,
and coupled to a  transverse field $h$. At its ends, {\itshape i.e.}, at $r=1$ and $r=N$, the chain is coupled to 
magnetic reservoirs which are kept  at zero temperature ($T=0$).
The Hamiltonian of the chain is given by 
\begin{align}
{\cal H}_\text{C}= & -\frac{J}{2}\sum_{r=1}^{N-1}\left(\left(1+\gamma\right)\sigma_{r}^{x}\sigma_{r+1}^{x}+\left(1-\gamma\right)\sigma_{r}^{y}\sigma_{r+1}^{y}\right)\nonumber \\
 & -h\sum_{r=1}^{N}\sigma_{r}^{z},\label{eq:Hamiltonian_C}
\end{align}
 where $\sigma_{r}^{x,y,z}$ are the Pauli matrices at  site $r$, and $\gamma$ controls the anisotropy. 
The total Hamiltonian is given by 
\begin{align}
{\cal H}=  {\cal H}_\text{C}+\sum_{l=\text{L},\text{R}}\left({\cal H}_{l}+{\cal H}_{\text{C-}l}\right),\label{eq:Hamiltonian}
\end{align}
where ${\cal H}_{l}$ and ${\cal H}_{\text{C-}l}$, with $l=\t{L,R}$, are respectively the Hamiltonians of the reservoirs and the system-reservoir coupling terms.  
In the following, we assume that the reservoirs possess bandwidths which are entirely determined by  magnetic potential $\mu_l$ ($l=\t{L,R}$) and which are much larger than the  energy scales that characterize the chain. 

\begin{figure}[t]
\begin{centering}
\hfill{}\includegraphics[width=0.75\columnwidth]{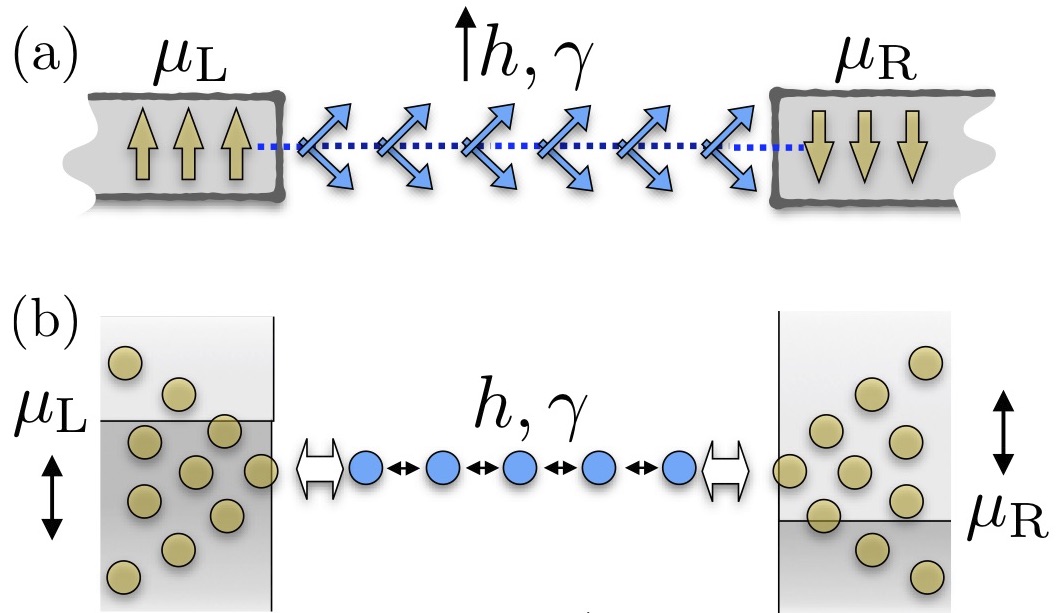}\hfill{}
\par\end{centering}
\centering{}\caption{\label{fig: Schematic picture of the model} (a) Schematic picture
of the XY-model spin chain in contact with magnetic reservoirs and
(b) the same system mapped to its fermionic representation, i.e. a
triplet superconducting chain of spinless fermions in contact with
fermionic reservoirs.}
\end{figure}

In the wide-band limit, results become independent of the  details of ${\cal H}_{l}$ and ${\cal H}_{\text{C-}l}$. For concreteness, we take the reservoirs to be isotropic $XY-$chains, i.e.
\begin{equation}
{\cal H}_{l}=-J_{l}\sum_{r_{l}\in\Omega_{l}}\left(\sigma_{r_{l}}^{x}\sigma_{r_{l}+1}^{x}+\sigma_{r_{l}}^{y}\sigma_{r_{l}+1}^{y}\right),
\end{equation}
with $l=\text{L},\text{R}$, and we have defined $\Omega_{\text{L}}\equiv\left\{ -\infty,\ldots,0\right\} $,
$\Omega_{\text{R}}\equiv\left\{ N+1,\ldots,\infty\right\} $.
Initially, the reservoirs are in an equilibrium Gibbs state, $\rho_l = e^{-\beta (\mathcal{H}_l - \mu_l M_l) }$, where $M_{l}=\sum_{r_{l}\in\Omega_{l}}\sigma_{r_{l}}^{z}$ is the reservoir magnetization 
(which is a good quantum number in the absence of system-reservoir coupling, {\itshape i.e.}, $\left[{\cal H}_{l},M_{l}\right]=0$).
The average value of $M_{l}$ is set by the magnetic potential $\mu_{l}$.
For finite $\mu_l$ these are non-Markovian reservoirs, with power-law decaying correlations,
and a set of gapless magnetic excitations within
an energy bandwidth $J_{l} \gg J, h$. 
The chain-reservoir coupling Hamiltonians
are 
\begin{equation}
{\cal H}_{\text{C-}l}=-J_{\text{C-}l}\left(\sigma_{\left(r_{l}\right)_{\text{C}}}^{x}\sigma_{\left(r\right)_{l}}^{x}+\sigma_{\left(r\right)_{l}}^{y}\sigma_{\left(r_{l}\right)_{\text{C}}}^{y}\right),
\end{equation}
with $\left(r_{\text{L}}\right)_{\text{C}}=1$, $\left(r\right)_{\text{L}}=0$,
$\left(r_{\text{R}}\right)_{\text{C}}=N$, and $\left(r\right)_{\text{R}}=N+1$.
A sketch of this system is shown in Fig.\ref{fig: Schematic picture of the model}(a).

The full Hamiltonian, ${\cal H}$, can be represented in terms of
fermions via the so-called Jordan-Wigner (JW) mapping\citep{Lieb1961},
$\sigma_{r}^{+}=e^{i\pi\sum_{r'=1}^{r-1}\hat{c}_{r'}^{\dagger}\hat{c}_{r'}}c_{r}^{\dagger}$,
where $\hat{c}_{r}^{\dagger}/\hat{c}_{r}$ creates/annihilates a spinless
fermion at site $r$. 
The JW-transformed system corresponds to a Kitaev chain\citep{Kitaev2001} in contact with two metallic reservoirs of spinless
fermions at chemical potentials $\mu_{\text{L},\text{R}}$, i.e.
\begin{align}
{\cal H}=-J\sum_{r=1}^{N-1}\left(\hat{c}_{r}^{\dagger}\hat{c}_{r+1}+\gamma\hat{c}_{r}^{\dagger}\hat{c}_{r+1}^{\dagger}+\text{h.c.}\right)-2h\sum_{r=1}^{N}\hat{c}_{r}^{\dagger}\hat{c}_{r} \nonumber \\
- \sum_{l=\text{L,R}} \left[ J_{\text{C-}l} \hat{c}_{\left(r_{l}\right)_{\text{C}}}^{\dagger}\hat{c}_{\left(r\right)_{l}}
+  J_{l} \sum_{r_{l}\in\Omega_{l}} \hat{c}_{r_{l}}^{\dagger}\hat{c}_{r_{l}+1} +\text{h.c.} \right],
\end{align}
where $J\gamma$ defines the superconducting coupling strength and $h$
plays the role a potential applied on the chain.
A sketch of this system is shown in Fig.\ref{fig: Schematic picture of the model}(b).
In equilibrium, topologically non-trivial phases of the Kitaev chain correspond to magnetically
ordered phases of the original XY-spin model, whereas the topologically trivial cases correspond to disordered phases.
With a magnetic bias, the transfer of spin excitations between the reservoirs was studied rather extensively (see, e.g., Refs.~\onlinecite{,PhysRevLett.90.167204,Nakata_2017}), also considering transport signatures of the topological phase in short junctions.\cite{hoffman2018superfluid,shen2020theory} Here, however, we will be mostly concerned with the bulk properties at $N\to \infty$, and after the reservoirs have been traced out. 

\subsection{Non-equilibrium Green's functions}

As the JW-transformed Hamiltonian is quadratic in its fermionic degrees of freedom, the non-equilibrium system admits an exact solution in terms of single-particle quantities. 
In the following, we employ the non-equilibrium Green's function formalism to compute correlation functions and related observables.   
The procedure is described in the Supplemental Material of Ref. [\onlinecite{Puel-Chesi-Kirchner-Ribeiro-2019}] and is briefly summarized here for convenience. 

We start by defining the  Nambu vector, $\hat{\boldsymbol{\Psi}}^{\dagger}=\left(\hat{c}_{1}^{\dagger},\ldots,\hat{c}_{N}^{\dagger},\hat{c}_{1},\ldots,\hat{c}_{N}\right)$, and the retarded, advanced, and Keldysh components of the Green's function, given by  
\begin{eqnarray}
\boldsymbol{G}^{R}_{i,j} (t-t') &=& - i \Theta\left(t-t'\right) \av{ \left\{   \hat{\boldsymbol{\Psi}}_i(t), \hat{\boldsymbol{\Psi}}_j^\dagger(t') \right\} }, \\
\boldsymbol{G}^{A}_{i,j} (t-t') &=&  i \Theta\left(t'-t\right) \av{ \left\{   \hat{\boldsymbol{\Psi}}_i(t), \hat{\boldsymbol{\Psi}}_j^\dagger(t') \right\} },  \\
\boldsymbol{G}^{K}_{i,j} (t-t') &=& - i \av{ \left[   \hat{\boldsymbol{\Psi}}_i(t), \hat{\boldsymbol{\Psi}}_j^\dagger(t') \right] }.
\end{eqnarray}
Using this notation, the Hamiltonians for the right, left reservoirs and for the chain are given by 
${\cal H}_{l}=\frac{1}{2}\hat{\boldsymbol{\Psi}}^{\dagger}\bs H_{ l}\hat{\boldsymbol{\Psi}}$, with $l=\text{L,R,C}$.
For the chain, $\bs H_{\text{C}}$ is a $2N\times2N$  Hermitian matrix respecting
 particle-hole symmetry, {\itshape i.e.}, $\bs S^{-1}\boldsymbol{H}_{\t C}^{T}\bs S=-\boldsymbol{H}_{\t C}$
where $\bs S=\tau^{x}\otimes\bs1_{N\times N}$ and $\tau^{x}$ interchanges
particle and hole spaces. Similar definitions apply to the degrees of freedom of the right and left reservoirs. 
The bare retarded and advanced Green's functions, in the absence of chain-reservoir couplings, are simply given by $ \boldsymbol{G}_{l,0}^{R/A}\left(\omega\right)=\left(\omega-\bs H_{l} \pm i \eta \right)^{-1}$. 
Re-writing the chain-reservoir coupling in the same notation, $ {\cal H}_{\text{C-}l} = \frac{1}{2} \left( \hat{\boldsymbol{\Psi}}^\dagger_{(l)} \bs T  \hat{\boldsymbol{\Psi}} + \hat{\boldsymbol{\Psi}}^\dagger \bs T^\dagger   \hat{\boldsymbol{\Psi}}_{(l)} \right)$. The self-energy of the chain, induced by tracing out the reservoirs, are $ \bs \Sigma^{R/A/K} = \sum_{l=\t{L,R} }\bs  \Sigma^{R/A/K}_l$ where
\begin{eqnarray}
\bs \Sigma^{R/A}_l (\omega) &=& \bs T^\dagger \boldsymbol{G}_{l,0}^{R/A}(\omega) \bs T , \\  
\boldsymbol{\Sigma}_{l}^{K}(\omega)&=&\left[\boldsymbol{\Sigma}_{l}^{R}(\omega)-\boldsymbol{\Sigma}_{l}^{A}(\omega)\right]\left[1-2 n_{\mathrm{F}, l}(\omega)\right],
\end{eqnarray}
are the contributions of reservoir $l$, which obey equilibrium fluctuation-dissipation relations, and where $n_{\text{F},l}(\omega) = (e^{\beta_l (\omega - \mu_l)}+1)^{-1}$ is the Fermi-function with chemical potential $\mu_l$ and inverse temperature $\beta_l$. 
The chain steady-state Green's functions are obtained from the Dyson's equation, 
\begin{eqnarray}
\boldsymbol{G}_{\text{C}}^{R/A}(\omega) &=& \left[ \boldsymbol{G}_{\text{C},0}^{R/A}(\omega)  - \boldsymbol{\Sigma}^{R/A}(\omega) \right]^{-1}, \\ \boldsymbol{G}_{\text{C}}^{K}(\omega) &=& \boldsymbol{G}_{\text{C}}^{R}(\omega)     \boldsymbol{\Sigma}^{K}(\omega)    \boldsymbol{G}_{\text{C}}^{A}(\omega).
\end{eqnarray}

As mentioned above, we consider the case where the bandwidths of the
reservoirs, $J_{l=\text{L},\text{R}}$ are much larger than the other
energy scales. In this wide band limit, the coupling to
reservoir $l$ is completely determined by the hybridization
energy scale $\Gamma_{l}=\pi J_{\text{C-}l}^{2}D_{l}$. Here, $D_{l}$ is the reservoir's constant-local density of states. 
In practice, the wide band limit yields a frequency independent retarded self-energy, $ \bs \Sigma^{R}_l = i ( \bs{\gamma}_{l} + \bar{\bs{\gamma}}_{l} ) $, which substantially simplifies subsequent calculations, with $\bs{\gamma}_{l}=\Gamma_{l}\ket{r_{l}}\bra{r_{l}}$ and $\bar{\bs{\gamma}}_{l}=\Gamma_{l}\ket{\bar{r}_{l}}\bra{\bar{r}_{l}}$, and 
where $\ket r$ and $\ket{\bar{r}}\equiv\bs S\ket r$ are single-particle and hole states. 

In this case, it is convenient to define the non-Hermitian single-particle operator
\begin{equation}
\bs K\equiv\bs H_{\t C}-i\sum_{l=\text{L,R}}\left(\bs{\gamma}_{l}+\bar{\bs{\gamma}}_{l}\right),\label{eq: K matrix}
\end{equation}
which we assume to be  diagonalizable, possessing right and left eigenvectors $\ket{\alpha}$
and $\bra{\tilde{\alpha}}$, and associated eigenvalues $\lambda_{\alpha}$.
In terms of these quantities, the retarded Green's function is given by 
\begin{equation}
\boldsymbol{G}^{R}\left(\omega\right)=\left(\omega-\boldsymbol{K}\right)^{-1}=\sum_{\alpha}\ket{\alpha}\left(\omega-\lambda_{\alpha}\right)^{-1}\bra{\tilde{\alpha}},
\end{equation}
and the Keldysh Green's function becomes
\begin{align}
& \bs G^{K}\left(\omega\right)  = -2i\sum_{l}\sum_{\alpha\beta}\ket{\alpha} \bra{\beta} \times \nonumber \\ 
&  \frac{\bra{\alpha'}\bs{\gamma}_{l}\ket{\beta'}\left[1-2n_{\t F,l}\left(\omega\right)\right]-\bra{\alpha'}\bar{\bs{\gamma}}_{l}\ket{\beta'}\left[1-2n_{\t F,l}\left(-\omega\right)\right]}{\left(\omega-\lambda_{\alpha}\right)\left(\omega-\bar{\lambda}_{\beta}\right)}. 
\end{align}

Steady-state observables can be obtained from the single-particle correlation function matrix, $\boldsymbol{\chi}\equiv\langle\hat{\boldsymbol{\Psi}}\hat{\boldsymbol{\Psi}}^{\dagger}\rangle$, 
which is obtained from the  Keldysh Green's function 
\begin{equation}
\boldsymbol{\chi}=\frac{1}{2}\left[i\int\frac{d\omega}{2\pi}\boldsymbol{G}^{K}\left(\omega\right)+\boldsymbol{1}\right].
\label{eq: single-particle density matrix}
\end{equation} 
The explicit form of $\boldsymbol{\chi}$ after performing the integration over frequencies is provided  in Eq.(\ref{eq:chi}). 
As the model is quadratic, $\boldsymbol{\chi}$ encodes all the information about the reduced density matrix of the chain, 
$\hat{\rho}_{\t C}= \tr_{\t{L,R}}[\hat{\rho}]$. This quantity can itself be expressed  as the exponential of a quadratic operator, i.e. 
$\hat{\rho}_{\t C}=\text{e}^{\hat{\Omega}_{\t C}}/Z$,
where $Z=\tr\left[\text{e}^{\hat{\Omega}_{\t C}}\right]$ and $\hat{\Omega}_{\t C}=\frac{1}{2}\hat{\boldsymbol{\Psi}}^{\dagger}\boldsymbol{\Omega}_{\t C}\hat{\boldsymbol{\Psi}}$
with $\boldsymbol{\Omega}_{\t C}$
being a $2N\times2N$ matrix respecting the particle-hole
symmetry conditions. $\hat{\Omega}_{\t C}$ is related to the single-particle density matrix via
\begin{equation}
\boldsymbol{\chi} = \left( \text{e}^{\boldsymbol{\Omega}_{\t C}} +\boldsymbol{1} \right)^{-1}.
\end{equation}
This relation allows the calculation of mean values of quadratic observables,  $\hat{O}=\frac{1}{2}\hat{\boldsymbol{\Psi}}^{\dagger}\boldsymbol{O}\hat{\boldsymbol{\Psi}}$, defined by the Hermitian and particle-hole symmetric matrix
$\boldsymbol{O}$, 
\begin{equation}
\left\langle \hat{O}\right\rangle =\Tr\left[\hat{\rho}_{\t C}\hat{O}\right]=-\frac{1}{2}\tr\left[\boldsymbol{O}\cdot\boldsymbol{\chi}\right],
\end{equation}
as well as all higher-order correlation functions. 

\section{Phase diagram\label{sec:Phase-diagram}}

This section discusses the non-equilibrium phase diagram of the model, as well as the excitations and associated occupation numbers  in the various different phases. 
To contextualize our findings,  the first two sub-sections are devoted to a brief description of the equilibrium properties of the $XY$-chain and a review of  the non-equilibrium Markovian limit. 

\subsection{Equilibrium phases}

The system in equilibrium is more conveniently studied without considering the couplings to the leads and assuming periodic boundary conditions.  
After performing the JW transformation, the Hamiltonian of the translation-invariant chain is diagonalized in the momentum representation by a suitable Bogoliubov transformation, {\itshape i.e.},   ${\cal H}_{C}=\sum_{k}\varepsilon_{k}(\hat{\gamma}_{k}^{\dagger}\hat{\gamma}_{k}-1/2)$,
where the operators  $(\hat{\gamma}_{k},\hat{\gamma}_{-k}^{\dagger})^{T}=e^{i\theta_{k}\sigma_{x}}(\hat{c}_{k},\hat{c}_{-k}^{\dagger})^{T}$ describe excitations of energy
\begin{equation}
\varepsilon_{k}=2 J\sqrt{\left(h/J+\cos k\right)^{2}+\left(\gamma \sin k\right)^{2}}, \label{eq:disperssion}
\end{equation} 
and $\sin\left(2\theta_{k}\right)=-2J\gamma\sin(k)/\varepsilon_{k}$.

The  ground state is characterized by a vanishing number of Bogoliubov excitation, {\itshape i.e.}, $n_k=0$ where 
\begin{equation}
n_{k}\equiv\langle\hat{\gamma}_{k}^{\dagger}\hat{\gamma}_{k}\rangle.\label{eq:occupationo number definition}
\end{equation}
For $\abs{h/J}<1$, the ground-state is topologically non-trivial with positive and negative anisotropies ($\gamma > 0$ or $<0$) corresponding to opposite signs of the topological invariant, separated by a critical gapless  state for $\gamma =0$. At $\abs{h/J}=1$, the spectral gap vanishes and the system transitions into a topologically trivial phase at large $h$. 

\begin{figure}[t]
\begin{centering}
\hfill{}\includegraphics[width=0.98\columnwidth]{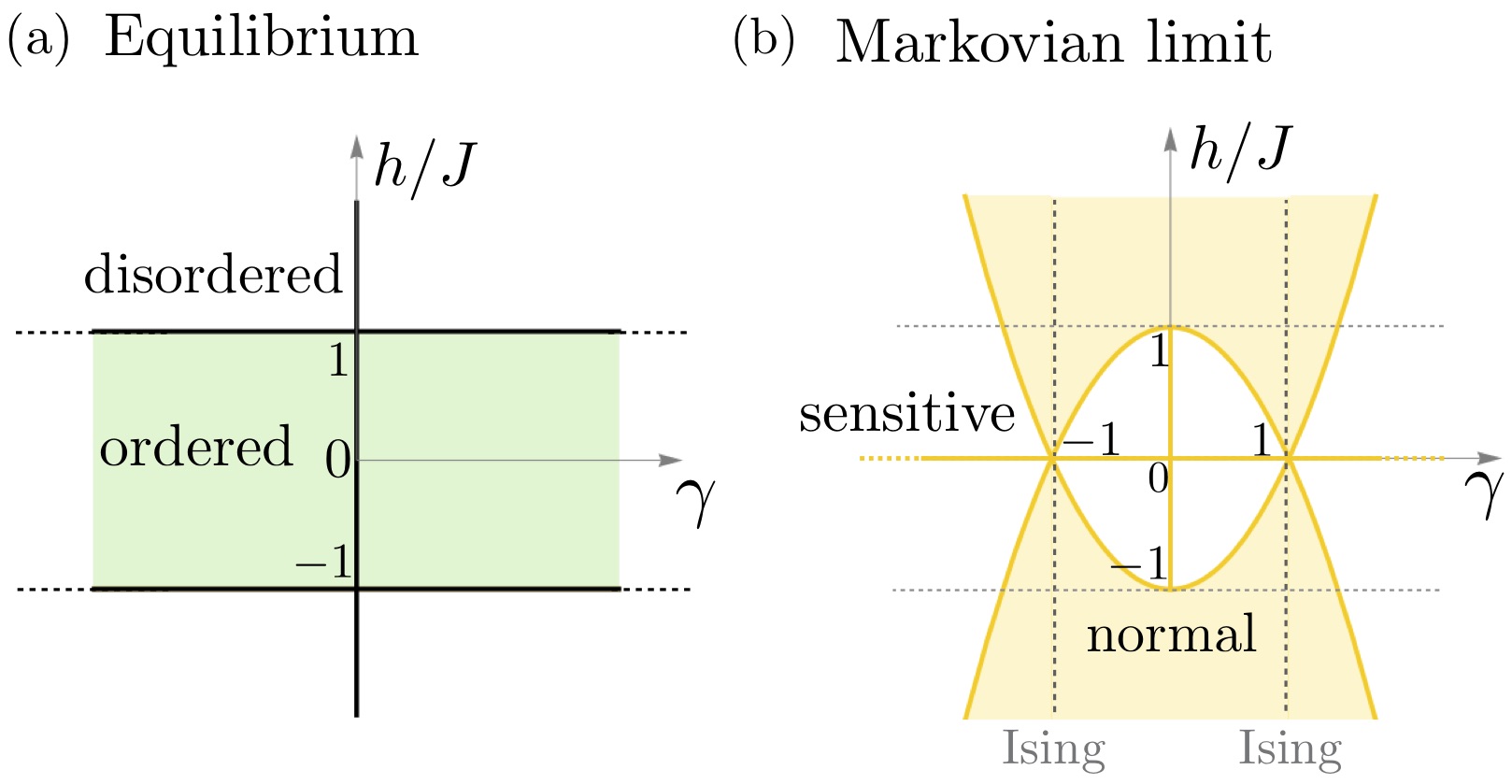}\hfill{}
\par\end{centering}
\centering{}\caption{\label{fig: equilibrium and markovian phase diagrams} (a) Phase diagram
$h/J\times\gamma$ of the XY-model in equilibrium. The black-thick
lines identify a gap closing. (b) Phase diagram of the out-of-equilibrium
XY-model according to the presence (sensitive) or absence (normal)
of long-range correlations in the Markovian limit. The vertical dashed
line at $\gamma=+1$ $(\gamma=-1)$ identifies the Ising model with
only XX(YY)-spin-coupling interactions.}
\end{figure}

A similar phase diagram is obtained in terms of the original spin degrees of freedom.  
Fig.\ \ref{fig: equilibrium and markovian phase diagrams}(a) illustrates the zero-temperature phase diagram of the equilibrium $XY$-model. 
For $\gamma > 0 $, the systems is magnetically ordered  along the $x$ direction under a weak transverse field, $h$, and possesses  a finite magnetization\citep{Barouch.71} $\phi \equiv \lim_{h_x \to 0} \lim_{L\to\infty} \frac{1}{N}\sum_r\av{\sigma^x_r} \neq 0$, where $h_x$ is a symmetry-breaking magnetic field along the $x$ direction. 
A negative anisotropy, {\itshape i.e.}, $\gamma<0$, yields a non-vanishing magnetization along the $y$ direction, whereas at $\gamma=0$ the system  is critical and isotropic for $\abs{h/J}<1$. As the ordered phases for $\gamma>0$ and $<0$ are equivalent to each other and related via a simple rotation, only $\gamma > 0$ is considered in the subsequent analysis. 
It is worth recalling that the special cases $\gamma=\pm 1$  correspond to the transverse field Ising model. 
A strong $h$ drives the magnetic phase through a second order phase transition into a phase of vanishing magnetization, {\itshape i.e.}, a phase with  $\phi=0$. 
Near the transition, for $\left|h/J\right|<1$, the magnetization behaves as $\phi \simeq \sqrt{ \frac{2}{1+|\gamma|} } \left\{ \gamma^2 \left[ 1 - \left( h/J \right)^2 \right] \right\}^{1/8}$ \cite{Barouch.71}. 

The computation of the order parameter, $\phi$, directly from the above definition is not possible via the JW mapping. Instead one considers the  
two-points correlation functions ($\alpha=x,y,z$):
\begin{equation}
\mathbb{C}_{r,r'}^{\alpha\alpha}=\langle\sigma_{r}^{\alpha}\sigma_{r'}^{\alpha}\rangle
-\langle\sigma_{r}^{\alpha}\rangle\langle\sigma_{r'}^{\alpha}\rangle.\label{eq: correlation definition}
\end{equation}
For disordered phases in equilibrium, these correlators are expected to show either exponential (EXP) or power-law (PL) decay depending on whether the system is gapped or gapless. In the ordered phase the system has long-range order (LRO) correlations, {\itshape e.g.}, for $\gamma>0$, 
\begin{equation}
\mathbb{C}_{r,r'}^{xx} \simeq A\text{e}^{-\left|r-r'\right|/\xi}+\phi^{2},\label{eq:correlation and correlation length}
\end{equation}
 where $\xi$ is the characteristic correlation length and $A$ a numeric coefficient. This expression allows one to obtain $\phi$ from the correlation function $\mathbb{C}_{r,r'}^{xx}$, which in turn can be computed in terms of a Toeplitz determinant \citep{lieb1961two,sachdev1996universal}. 
 In equilibrium,  all correlation functions, except  $\mathbb{C}_{r,r'}^{xx}$,  either vanish or decay exponentially with $\abs{r-r'}$.
 
 For an open system connected to demagnetized baths, {\itshape i.e.}, $\mu_{L,R} =0 $,   the same equilibrium bulk properties as those for the closed system are found.
 It is natural to expect that bulk properties pertain for distances greater than $\xi$ away from the leads. Our calculation of fermionic observables follows that described in  Sec.\ref{sec:Model-and-Method}. The calculation of the spin-spin  correlation functions are similar to those for the translation-invariant system and is given in Appendix \ref{app:Method-detailed} in terms of the single-particle correlation matrix $\bs \chi$.

\subsection{Nonequilibrium phases in the Markovian limit}

The non-equilibrium features of the XY-model with Markovian reservoirs  were first reported in Refs. [\onlinecite{Prosen2008,Banchi2013}]. This limit can be recovered from the present model by taking $\left|\mu_{\text{R}}\right|$
or $\left|\mu_{\text{L}}\right|\rightarrow \infty$\citep{Ribeiro2014e}.
The steady-state phase diagram in that limit possesses two distinct phases characterized respectively by an exponential decay of all correlation functions with distance and by an algebraic decay of $\mathbb{C}^{zz}_{r,r'}$ concomitant with a strong sensitivity to small variations of some control parameters\citep{Prosen2008}.
Fig. \ref{fig: equilibrium and markovian phase diagrams}(b) depicts the phase diagram in this Markovian limit and with its sensitive (white) and  normal  (light-yellow) phases. The dark yellow lines mark critical phase boundaries. 

\begin{figure}[t]
\begin{centering}
\hfill{}\includegraphics[width=0.8\columnwidth]{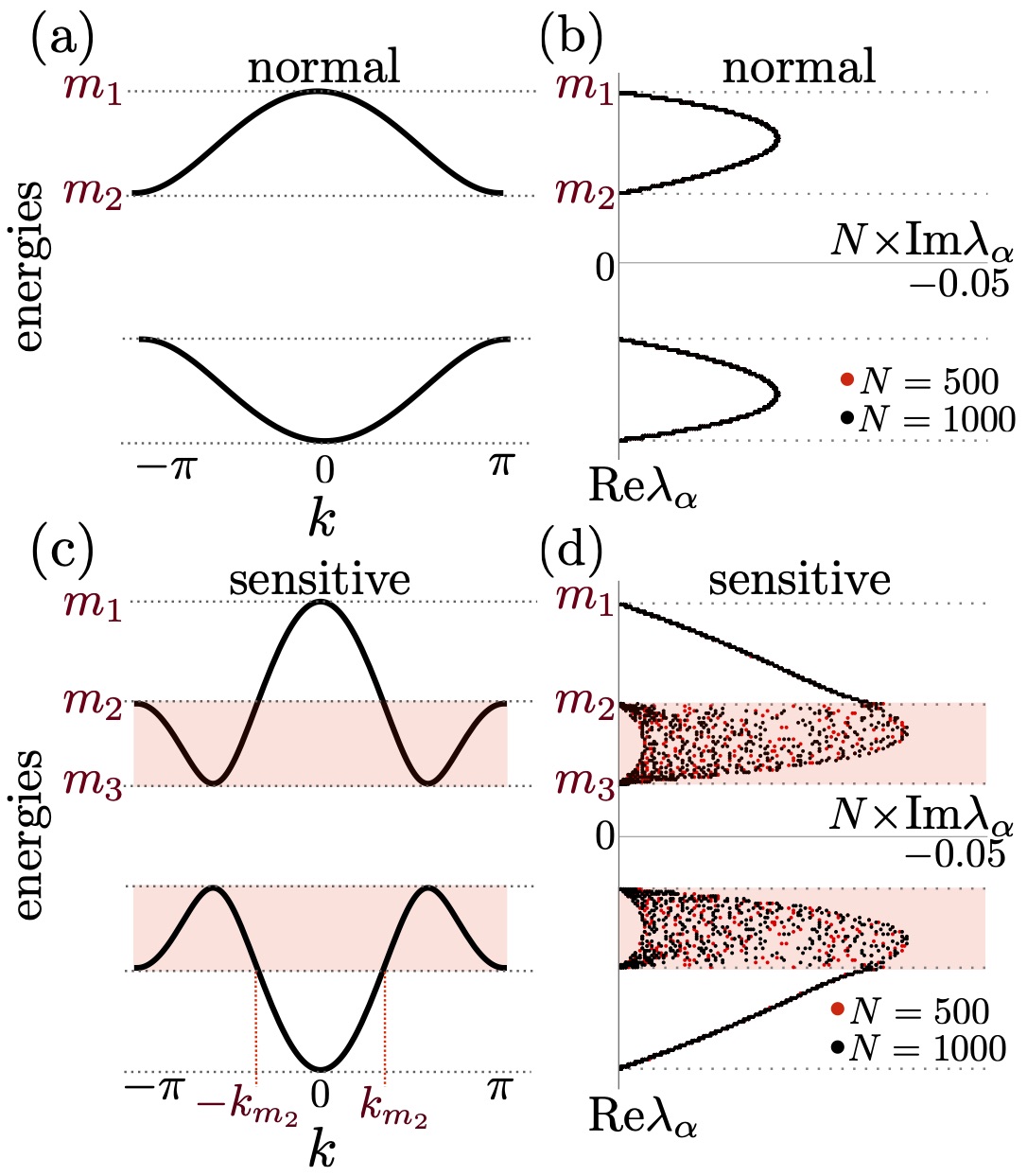}\hfill{}
\par\end{centering}
\centering{}\caption{\label{fig: band spectrum and rapidities} Quasiparticle dispersion relation 
of the equilibrium XY-model in the normal (a) and sensitive (c) regions, with $\{\gamma,h/J\}=\{1.0,0.2\}$ and $\{\gamma,h/J\}=\{0.5,0.2\}$ respectively, 
which leads to $\{m_1, m_2, m_3\} = \{2.4,1.6,0.97\}$.
$\pm k_{m_{2}}$ are the two momenta with energy $m_{2}$. 
(b) and (d) depict the eigenvalues $\lambda_{\alpha}$ of the non-Hermitian single-particle
operator $\boldsymbol{K}$ in the complex plane. }
\end{figure}

The quasiparticle dispersion relation, given by Eq.(\ref{eq:disperssion}), is shown in \ref{fig: band spectrum and rapidities}(a) and (c).
The algebraic correlations are associated to the presence of an inflection point in the quasiparticle dispersion which appears for $h/J\leq|1-\gamma^{2}|$. 
In the normal region, the extrema of the energy are $m_1 = \varepsilon_{k=0} = 2J\left|1+h/J\right|$ and $m_2 = \varepsilon_{k=\pi}=2J\left|1-h/J\right|$, while for the sensitive region $m_3 = 2J \left|\gamma\right|\sqrt{1+\left(h/J\right)^{2}\left(\gamma^{2}-1\right)^{-1}}$ becomes a global extremum. 
Figs. \ref{fig: band spectrum and rapidities}(b) and \ref{fig: band spectrum and rapidities}(d) show the spectrum of the non-Hermitian single-particle operator $\boldsymbol{K}$ (see Eq. (\ref{eq: K matrix})) for both normal and sensitive phases. 
It turns out that the imaginary part of the eigenvalues scales with the inverse system size, $\text{Im}\lambda_{\alpha}\propto N^{-1}$. \citep{Medvedyeva2014}
This is a reflection of the fact that for the chain degrees of freedom, the dissipative effects of the boundary become less important with increasing  system size. 
A key feature of the sensitive region is that, for energies ({\itshape i.e.} $\re(\lambda_\alpha) $) where four momenta can propagate, the spectrum does not converge to a line with increasing system size, but becomes scattered within a finite area \citep{Prosen2008}. 
These effects are independent of the Markovian nature of the reservoirs and remain for the non-Markovian case as the operator $\boldsymbol{K}$ does not depend on the chemical potential of the leads. 
Thus, as explicitly shown below, the normal-sensitive transition, reported in Refs. [\onlinecite{Prosen2008,Banchi2013}] for the Markovian case, also occurs for finite values of $\mu_{\text{L}}$ and $\mu_{\text{R}}$.

\subsection{Non-equilibrium phase diagram - energy current}

We now present the phase diagram of non-equilibrium $XY$ model and show that the energy current passing through the chain can be used to discriminate between the different phases. 

Conservation of energy implies that the steady-state energy current  is equal across any cross section along the chain and can be
obtained from $\bs{\chi}$ as 
\begin{equation}
\mathcal{J}_{e}=-\frac{1}{2}\,\tr\left[\boldsymbol{J}_{r}\cdot\boldsymbol{\chi}\right],\label{eq: current of energy}
\end{equation}
where $\boldsymbol{J}_{r}$ is the single-particle current operator at link $(r,r+1)$ which is explicitly
given in Appendix \ref{app:Method-detailed}. 

\begin{figure}[t]
\begin{centering}
\hfill{}\includegraphics[width=0.99\columnwidth]{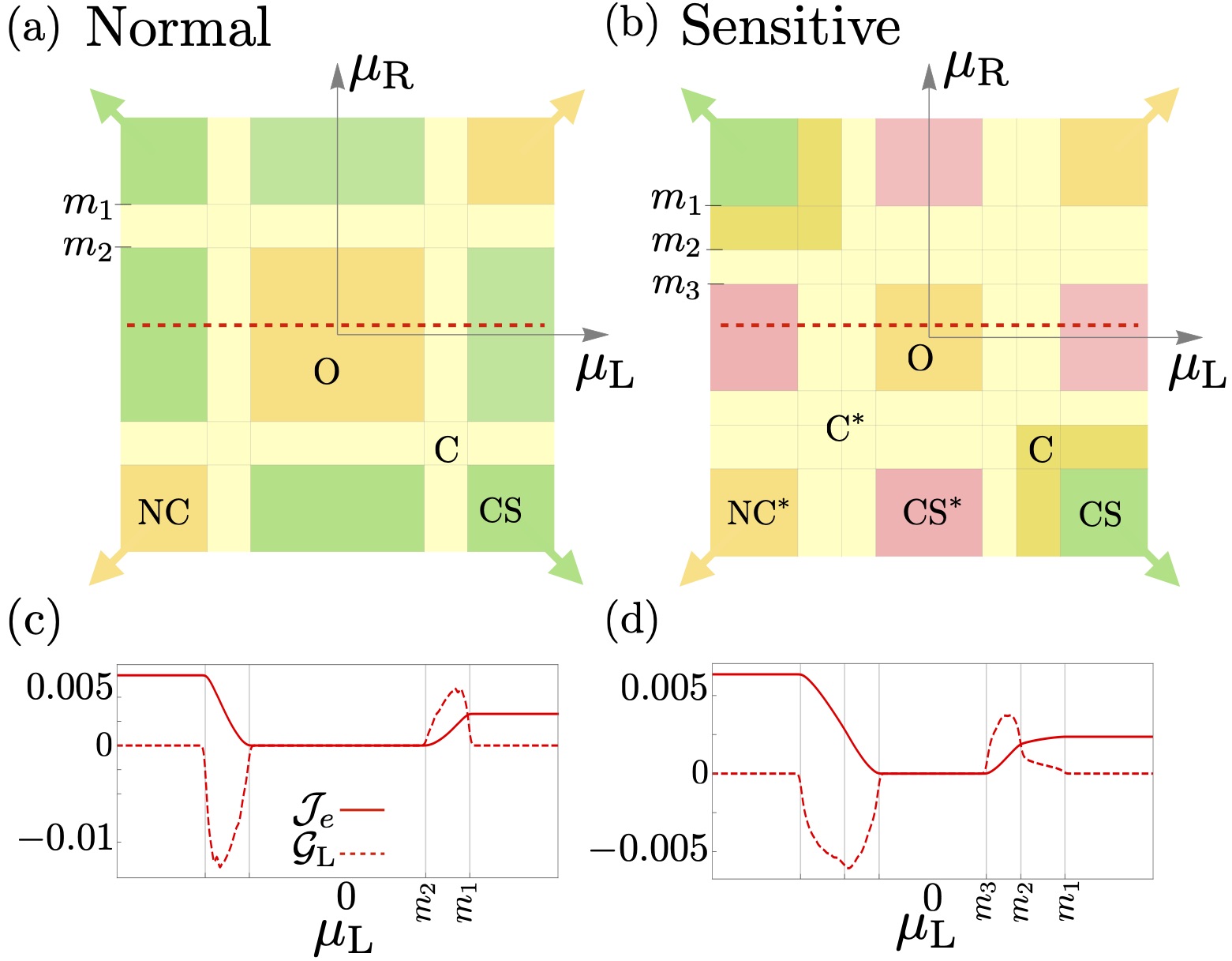}\hfill{}
\par\end{centering}
\centering{}\caption{\label{fig: non-Markovian phase diagram} 
Non-Markovian phase diagram $\mu_{L}\times\mu_{R}$ of two illustrative settings inside the 
(a) normal and (b) sensitive regions, following the same set of parameters in Fig. \ref{fig: band spectrum and rapidities}.
The phases were defined as ordered (O), conducting (C), conducting
saturated (CS), and non-conducting (NC).
In the sensitive case, the phases which acquire a noise in the occupation number are signed with a star $^*$.
The arrows at the corners indicate the Markovian limit, i.e. $\left|\mu_{\text{R}}\right|$
and $\left|\mu_{\text{L}}\right|\rightarrow\pm\infty$. (c) and (d)
show the current of energy ($\mathcal{J}_{e}$) and the conductance
(${\cal G}_{\t L}\equiv\pd_{\mu_{\t L}}\mathcal{J}_{e}$) computed
across the red-dotted lines drawn on the phase diagrams (a) and (b),
respectively.}
\end{figure}

We have previously discussed the  steady-state energy current in a non-Markovian setting  for the particular case of the transverse field Ising model, {\itshape i.e.}, $\left|\gamma\right|=1$, in Ref.  [\onlinecite{Puel-Chesi-Kirchner-Ribeiro-2019}].
 The non-equilibrium phase diagram, as a function of $\mu_{\text{L}}$ and $\mu_{\text{R}}$, in the normal phase, is qualitatively similar to the Ising  case and is reproduced in Fig. \ref{fig: non-Markovian phase diagram}(a).
Two of the phases which arise near $\mu_{\text{L}}=\mu_{\text{R}}$, do not support energy transport,{\itshape i.e.}, $\mathcal{J}_{e}=0$: the ordered phase (O) and the non-conducting phases (NC). 
Other phases may be further characterized in terms of their energy conductance, {\itshape i.e.}, ${\cal G}_{\t L}\equiv\pd_{\mu_{\t L}}\mathcal{J}_{e}$ and ${\cal G}_{\t R}\equiv\pd_{\mu_{\t R}}\mathcal{J}_{e}$. 
We refer to current-saturated (CS) the phases with $\mathcal{J}_{e}\neq0$ and ${\cal G}_{\t R}={\cal G}_{\t R}=0$. They arise when one of the reservoirs chemical potentials is larger than $m_{1}$ while the other lies inside the quasiparticle excitation gap. 
The conducting phase (C) is characterized by a non-zero conductance, {\itshape i.e.},
${\cal G}_{\t L}\neq0$ and/or ${\cal G}_{\t R}\neq0$, arising whenever
at least one of the chemical potentials lies within the quasiparticle
excitations band, {\itshape i.e.}, $\abs{\mu_{\text{L}}}$ and/or $\abs{\mu_{\text{R}}}\in(m_{1},m_{2})$.

Fig. \ref{fig: non-Markovian phase diagram}(b) depicts the phase diagram for a generic $XY$ chain. Besides the phases found for $\gamma=1$, an analysis of the occupation numbers (see next subsection) shows that some regions acquire a noise-like behavior. These phases, similar to the sensitive regions of the Markovian case, are labelled NC$^*$, CS$^*$, and C$^*$.

In Figs. \ref{fig: non-Markovian phase diagram}(c) and \ref{fig: non-Markovian phase diagram}(d)
we show the current $\mathcal{J}_{e}$ and conductance ${\cal G}_{\t L}$
for a fixed $\mu_{\text{R}}$ represented by the red-dashed lines in the phase diagrams. 
In the sensitive region, the C phase is crossed by the transition line at $\mu_L=m_{2}$, where the conductance becomes non-analytic.
It is worth noting that the noise-like behavior found in the occupation numbers does not appear in the current of energy.

In terms of the JW fermions, the present analysis is similar to that of a transport across a tight binding model in the sense that when the chemical potentials cross  the dispersion relation non-analytic properties of the current appears. However, the existence of anomalous terms in the fermionic Hamiltonian inhibits a closer comparison with charge transport.

\subsection{Occupation numbers\label{subsec:Occupation-number}}

In the current carrying steady-state regime of a Fermi gas, fluctuations in the number of particles were shown to be intimately related to the entanglement entropy of a subsystem \citep{PhysRevLett.102.100502,PhysRevB.82.012405,PhysRevB.83.161408,PhysRevB.85.035409}. 
In analogy, the occupation number of the Bogoliubov excitations, given by Eq.(\ref{eq:occupationo number definition}), can be used to describe the properties of the asymptotic steady-state away from the boundaries. 
In the open system setting, $n_{k}$ can be approximated by numerically computing the Fourier transform 
$\bs\chi_k =  \sum_{r \in \Omega }  e^{-i k (r-r_0)} \bs\chi_{r,r_0} $, where $\Omega=\left\{r: L/4<r <3L/4 \right\}$ and $r_0 = L/2$, followed by a Bogoliubov transformation.  
In equilibrium, $\mu_L=\mu_R=0$, $n_{k}\simeq 0$ as expected, while in a generic out-of-equilibrium situation $n_{k}\neq 0$.

For the Ising model, it was shown that in the CS and NC phases the $n_{k}$ is a continuous function
of $k$, while in the C phase it has discontinuities depending on
the reservoir's chemical potentials \citep{Puel-Chesi-Kirchner-Ribeiro-2019}.
These discontinuities happen at the momenta $\pm k_{l=\text{L,R}}$
where the chemical potential $\mu_{l=\text{L,R}}$ cross the dispersion relation, see Fig. \ref{fig:occupation_numbers}(a). 
These results extent straightforwardly to the  XY-model in the normal region, see Fig. \ref{fig:occupation_numbers}(c)-(f). 
Within the O phase the system behaves as in equilibrium, i.e. $n_{k}\simeq 0$.

In the sensitive region there may be two absolute values of momenta,
labelled $\pm k_{l=\text{L,R}}$ and $\pm k'_{l=\text{L,R}}$, for which each chemical potential crosses the dispersion relation, as illustrated in Fig. \ref{fig:occupation_numbers}(b).
Interestingly, we find that $n_{k}$ has an intrinsic noise in the sensitive region, see
Figs. \ref{fig:occupation_numbers}(g)-(j).  
The noise appears in phases  NC$^*$, C$^*$, and CS$^*$, for $\left|k\right|>k_{m_{2}}$,  where $\varepsilon_{k_{m_2}} = m_{2}$ (see Fig. \ref{fig: band spectrum and rapidities}(c)).
In Appendix \ref{app: excitation number} we check that the magnitude of the noise in $n_k$ does not diminishes with increasing system sizes.
Curiously, the noise vanishes along the line $\mu_{\text{L}}=-\mu_{\text{R}}$, as well as within phases C and CS crossed by this line, as shown in Figs. \ref{fig:occupation_numbers}(k) and (l), and studied in detail in Appendix \ref{app: excitation number}. 

Note that $n_{k}$ is asymmetric upon changing $k\to-k$ for all conducting phases as required to maintain a net energy flow through the chain, as $\varepsilon(k)=\varepsilon(-k)$. Figs. \ref{fig:occupation_numbers}(c)-\ref{fig:occupation_numbers}(l)
illustrates this feature by showing a larger value of the hybridization, that yield a larger current of energy and consequently to a more
asymmetric $n_{k}$.
\begin{center}
\begin{figure}[t]
\includegraphics[width=0.98\columnwidth]{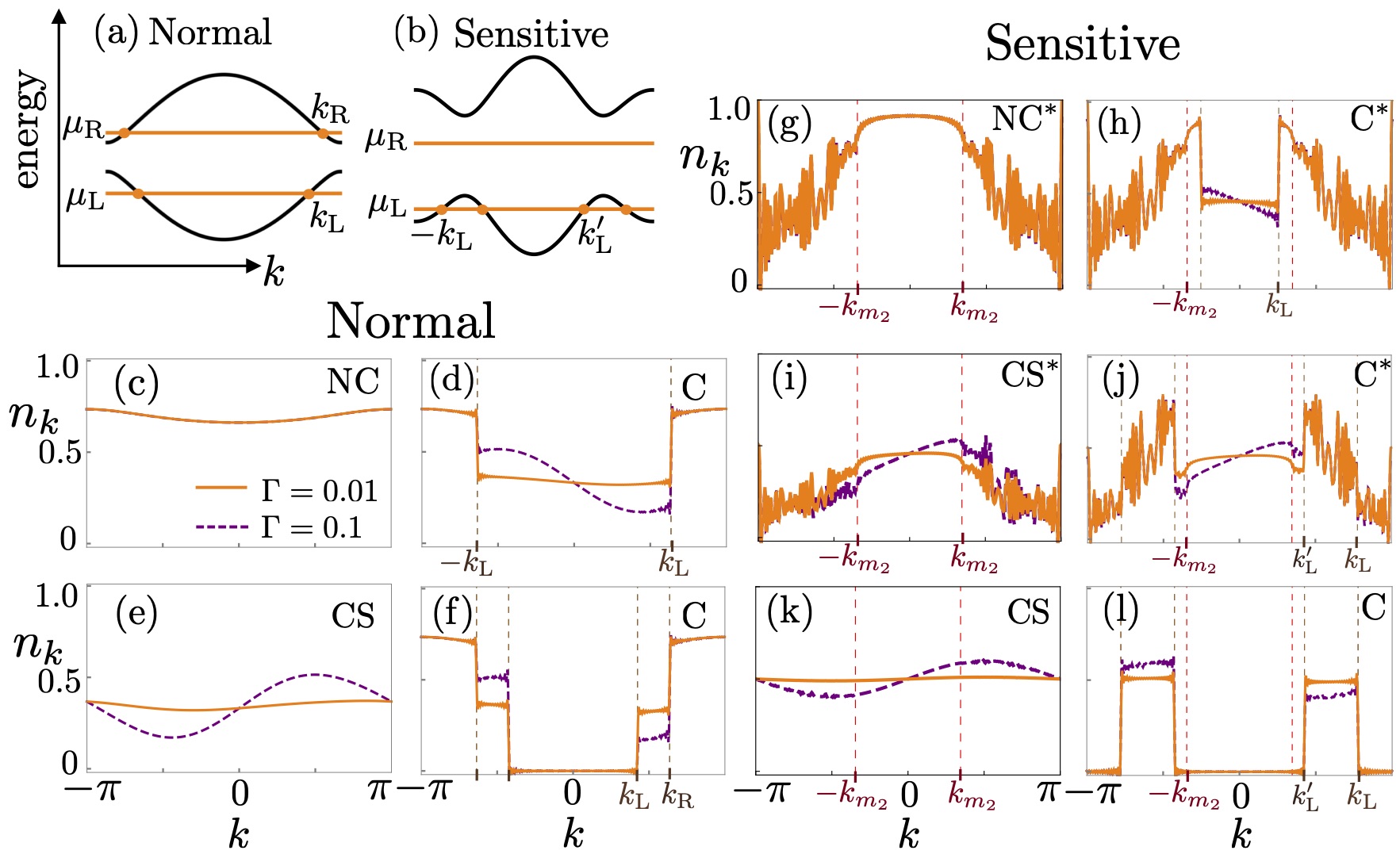}\caption{\label{fig:occupation_numbers} 
Distribution of occupations.
(a) and (b) indicate the momenta
$k_{l=\text{L,R}}$ at which the reservoir's chemical potentials $\mu_{l=\text{L,R}}$
cross the dispersion relation. (c)-(f) illustrate the
excitation number $n_{k}$ at different phases inside the normal region.
(g)-(l) shows $n_{k}$ in the sensitive region. $k_{m_2}$ is such that $\varepsilon_{k_{m_2}} = m_{2}$, with $m_{2}$ given in Fig. 
\ref{fig: band spectrum and rapidities}(c).  }
\end{figure}
\par\end{center}

\section{Correlation functions\label{sec: mixed-order phase transition}}

\begin{figure}
\begin{centering}
\hfill{}\includegraphics[width=0.95\columnwidth]{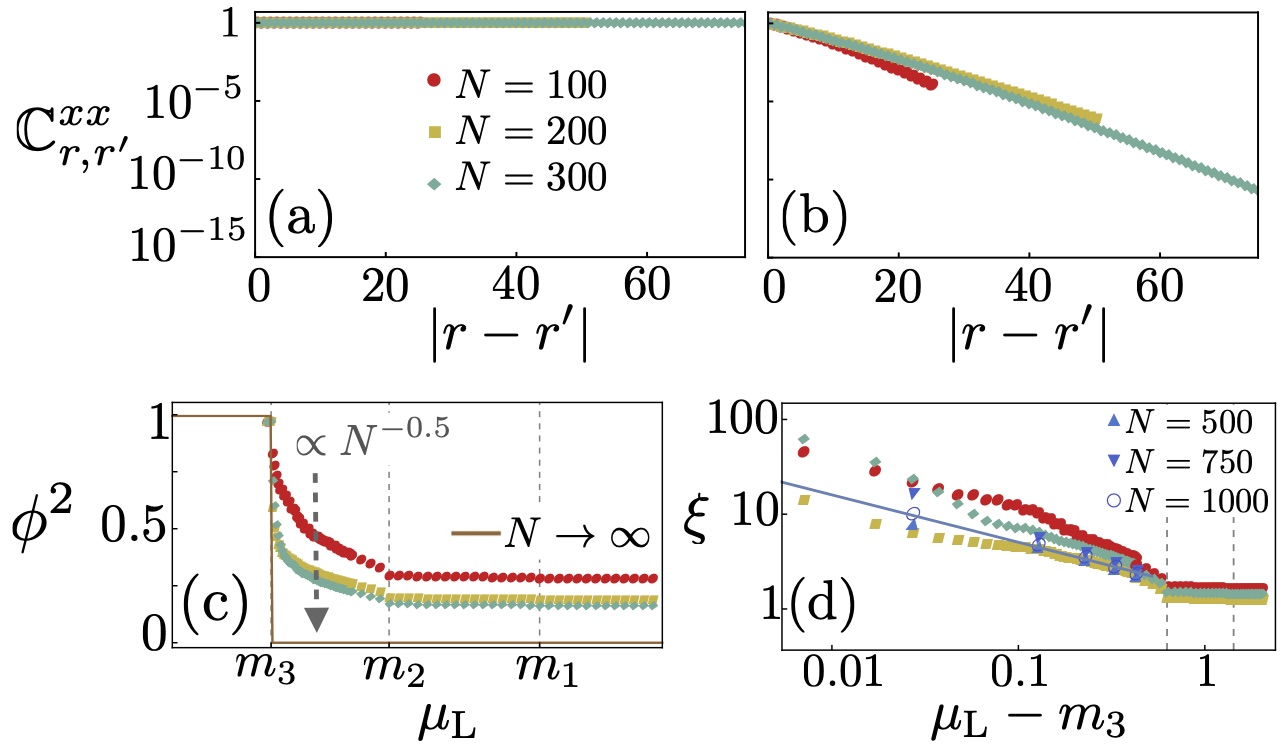}\hfill{}
\par\end{centering}
\centering{}\caption{\label{fig: xx correlations} (a) and (b) illustrate the correlation
$\mathbb{C}_{r,r'}^{xx}$ behavior in the nonequilibrium phases of Fig. \ref{fig: non-Markovian phase diagram}(b), with fixed $\mu_{\text{R}} = 0$. 
Long-range order only appears in the ordered phase (O), as exemplified in panel (a) at the point $|\mu_{\text{L}}| = 0.3 < m_3$, 
while a typical exponential
decay appears in all other phases, as exemplified in panel (b) for the C$^*$ phase at $m_3 < |\mu_{\text{L}}|=1.3 < m_2$.
(c) and (d) illustrate
the mixed-order behavior with a discontinuous order parameter $\phi$
and diverging correlation length $\xi$, respectively, in the thermodynamic
limit. The straight line in (d) is a guide to the critical exponent $\nu=1/2$. All panels share the same legends as in (a).
}
\end{figure}

We now consider in more detail the properties of the spin correlation functions, defined in Eq.~(\ref{eq: correlation definition}). For $\mathbb{C}_{r,r'}^{xx}$, the generic asympototic dependence was already given in Eq.~(\ref{eq:occupationo number definition}) and is able to signal the presence of long-range order when $\phi\neq 0$. We give in Fig.~\ref{fig: xx correlations}(a) a numerical example of this case. On the other hand, when $\phi=0$ we can extract the correlation length $\xi$ from the exponential decay of $\mathbb{C}_{r,r'}^{xx}$, see Fig.~\ref{fig: xx correlations}(b).  
Table~\ref{tb: table correlations} shows a summary of the asymptotic dependence of $\mathbb{C}_{r,r'}^{xx}$ in the different phases.

\begin{table}
\begin{tabular}{|c|c|c|}
\hline 
Phase & $\mathbb{C}_{r,r'}^{xx}\left(\gamma>0\right)$ & $\mathbb{C}_{r,r'}^{zz}$\tabularnewline
\hline 
O & LRO & EXP\tabularnewline
\hline 
C/C$^{*}$ & EXP & PL/PL\tabularnewline
\hline 
CS/CS$^{*}$ & EXP & EXP/PL\tabularnewline
\hline 
NC/NC$^{*}$ & EXP & EXP/PL\tabularnewline
\hline 
\end{tabular}

\caption{Classification of each phase according to the asymptotic behavior of the correlation functions; the possibilities are exponential (EXP) or power law (PL) decay, and long-range order (LRO).}

\label{tb: table correlations}

\end{table}

The typical dependence of $\phi$ and $\xi$ on the chemical potential of the reservoirs is illustrated in panels (c) and (d) of Fig.~\ref{fig: xx correlations}, showing the remarkable property of a discontinuity in $\phi$ at the critical point (after extrapolation to the thermodynamic limit) accompanied by a diverging correlation length $\xi$. 
Further below we will elaborate on the mixed-order transition in more detail, by also providing an analytical description clarifying its origin. 

\begin{figure}[t]
\begin{centering}
\hfill{}\includegraphics[width=0.99\columnwidth]{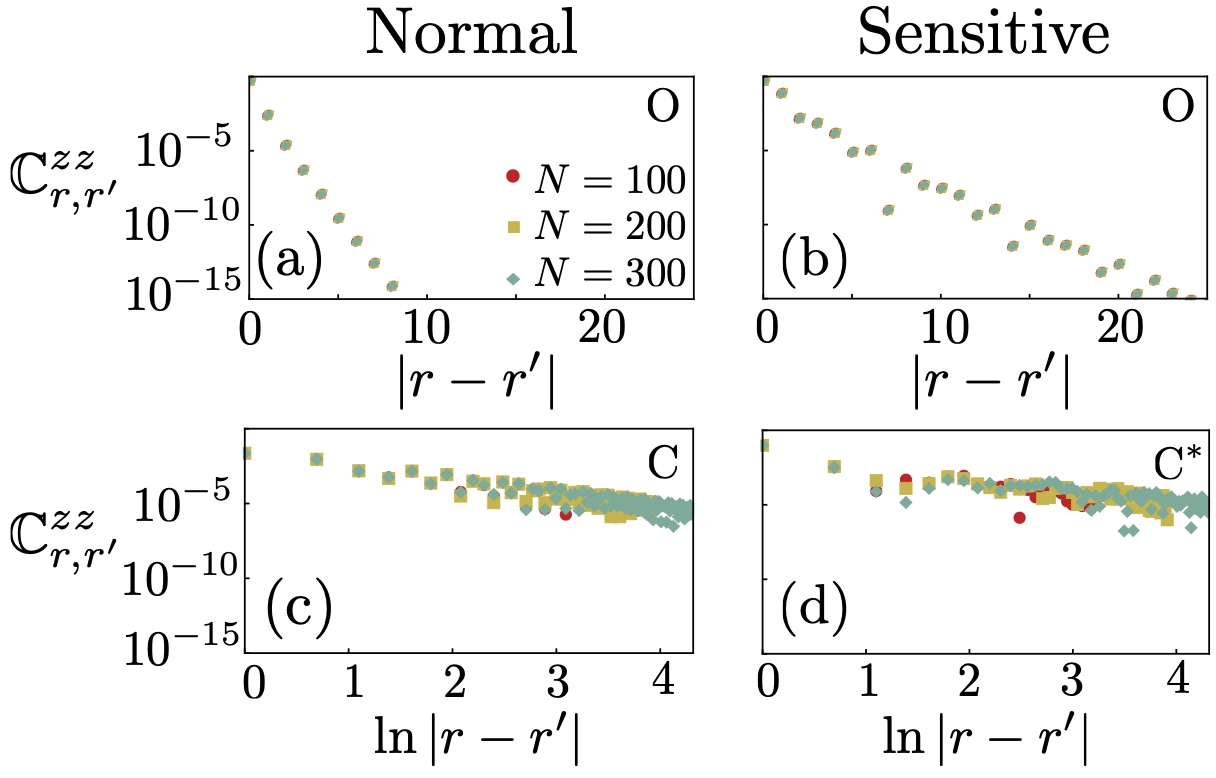}\hfill{}
\par\end{centering}
\centering{}\caption{\label{fig: z-correlation function} Correlations $\mathbb{C}_{r,r'}^{zz}$
for the normal (left column) and sensitive (right column) regions, see Fig. \ref{fig: non-Markovian phase diagram}.
We have set (a) and (b) at $\mu_{\text{L}} = \mu_{\text{R}} = 0$, (c) at $m_2 < |\{\mu_{\text{L}},\mu_{\text{R}}\}=\{2.1,-2.1\}| < m_1$, and (d) at $m_3 < |\{\mu_{\text{L}},\mu_{\text{R}}\}=\{1.3,-1.3\} | < m_2$.
These
results are summarized in Table \ref{tb: table correlations}. All panels share the same legends as in (a). 
}
\end{figure}

As shown in Table~\ref{tb: table correlations}, the behavior of $\mathbb{C}_{r,r'}^{xx}$ is not affected by the transition to the sensitive region. 
However, earlier studies showed how, in the Markovian limit, the CS and NC phases are characterized by a transition, from short- to long-range correlations, when entering the sensitive region. This behavior is reflected by a transition from exponential to power-law decay in $\mathbb{C}_{r,r'}^{zz}$ \cite{Prosen2008,_unkovi__2010}.
We observe similar results in the present case, with both non-conducting and saturated phases (CS and NC) showing exponentially-decaying correlations in the normal region and power-law decay in the sensitive one (CS$^*$ and NC$^*$).
Extending the analysis to the highly non-Markovian setting, we find that long-range correlations also appear in the conducting phase, with a power-law decay in both normal (C) and sensitive (C$^*$) regions, see Figs.~\ref{fig: z-correlation function}(c) and \ref{fig: z-correlation function}(d).
These results show that all phases with a noisy excitation number distribution possess power-law decaying $zz$ correlations.
The asymptotic behavior of $\mathbb{C}_{r,r'}^{zz}$ in the various phases is summarized in  Table~\ref{tb: table correlations}.

\subsection{Mixed-order phase transition}

As aforementioned, for $\gamma>1$, the magnetization along the $x$ direction, $\phi$, is a good order parameter for the broken-symmetry equilibrium phase.   
In the open system $\phi$ can still be used as order parameter.
However, by changing the chemical potential of the, say, left reservoir, $\phi$ drops to zero discontinuously as soon as the system reaches the disordered phase. 
Interestingly, this transition shows a mixed-order behavior where the discontinuity of $\phi$ is accompanied by a divergence of the correlation length \citep{Puel-Chesi-Kirchner-Ribeiro-2019}. 
The divergence occurs as $\xi\propto\left|\mu_{\text{L}}\pm m_{i}\right|^{-\nu}$ when approaching the critical point from the disordered phase ($i=2,3$ depending on the values of $\gamma$ and $h$). The critical exponent is $\nu=1/2$, except for special values of the parameters (see Appendix~\ref{app:analytical-approach-mixed-order}).

We have numerically verified that this behavior survives away from $\gamma=1$, with the same type of dependence of the order parameter and correlation length. In particular, the mixed-order transition with $\nu=1/2$ is also present in the sensitive region, e.g., at the transitions between O and C$^*$ phases in the phase diagram of Fig. \ref{fig: non-Markovian phase diagram}(b).
We show in panels (c) and (d) of Fig. \ref{fig: xx correlations} an example of the numerical analysis of order parameter and correlation length in the sensitive region. The main difference compared to the normal region is that here finite size effects are much stronger, which is consistent with the sensitive dependence on $N$ of the the rapidity spectrum and occupation numbers, see Figs.~\ref{fig: band spectrum and rapidities} and \ref{fig:occupation_numbers}. 

To derive the value of the critical exponent $\nu$, we consider the explicit form of the correlation function in terms of a Toepliz determinant:\citep{lieb1961two,sachdev1996universal}
\begin{equation}\label{Toepliz}
\left\langle \sigma_{r}^{x}\sigma_{r+n}^{x}\right\rangle =\left\vert \begin{array}{ccccc}
D_{0} & D_{-1} & . & . & D_{-n+1}\\
D_{1} & D_{0} & . & . & .\\
. & . & . & . & .\\
. & . & . & D_{0} & D_{-1}\\
D_{n-1} & . & . & D_{1} & D_{0}
\end{array}\right\vert ,
\end{equation}
where $D_{n}$ is given by:
\begin{equation}\label{eq:Dn}
D_{n}=\int\frac{dk}{2\pi}e^{-ink}\sqrt{\frac{1-\left(h/J\right)e^{ik}}{1-\left(h/J\right)e^{-ik}}}\left(1-n_{k}-n_{-k}\right).
\end{equation}
The asymptotic dependence can be obtained from Szego's lemma, leading to the following expression for the correlation length:
\begin{equation}\label{xi_Szego}
\xi^{-1} = -\frac{1}{2\pi}\int_{0}^{2\pi}\log\left[1-n_{k}-n_{-k}\right]dk.
\end{equation}
Here, the difference from the standard treatment of the transverse-field Ising chain\citep{lieb1961two,sachdev1996universal} is simply that the occupation numbers $n_k$ are kept generic, thus are allowed to assume any non-equilibrium distribution induced by the external reservoirs. For example,  Eq.~(\ref{xi_Szego}) takes into account that in general $n_{k} \neq n_{-k}$, as shown by Fig.~\ref{fig:occupation_numbers} with a large hybridization energy. By substituting the Fermi distribution, Eqs.~(\ref{eq:Dn}) and (\ref{xi_Szego}) recover the known equilibrium expressions at finite temperature.\cite{sachdev1996universal,Sachdev-2011} 

\begin{figure}[t]
\centering{}\includegraphics[width=0.98\columnwidth]{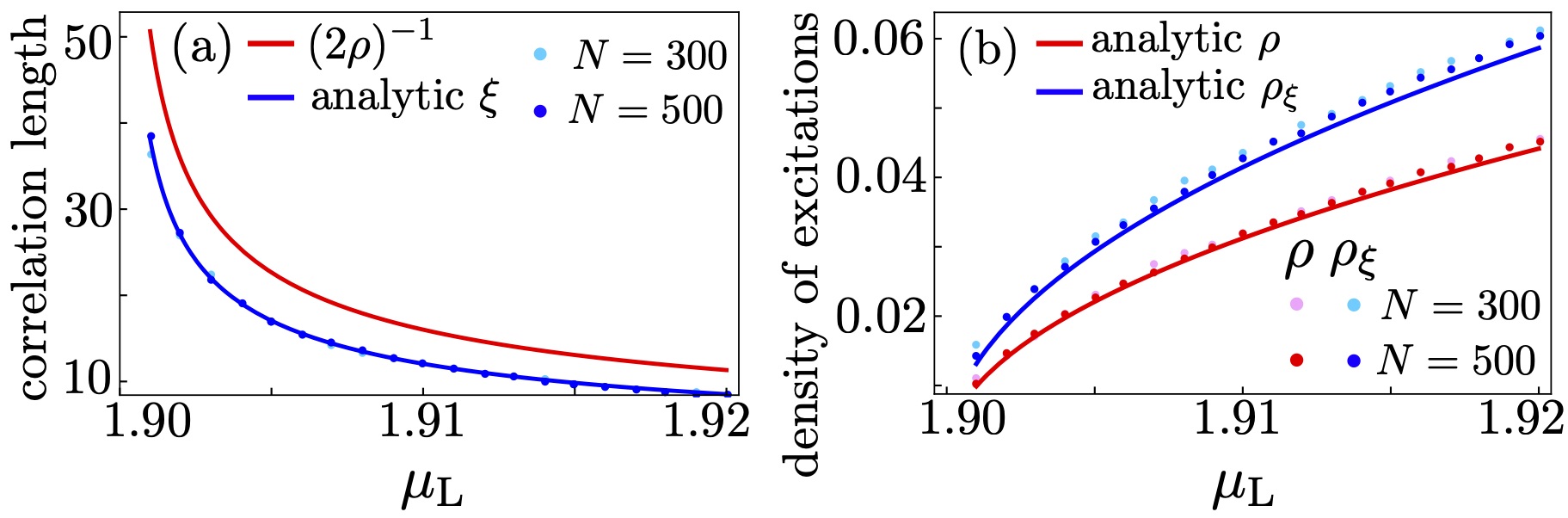}\caption{\label{fig: occupation number density} 
(a): The correlation length of the Ising model ($\gamma=1$), computed at weak transverse field ($h/J=-0.05$) and approaching the critical point $m_{1}=1.9$ from the disordered phase, with $\mu_{\text{R}} = 0$. 
The numerical results (dots) agree well with the blue (lower) curve, given by Eq.~(\ref{xi_divergence_accurate}). The red (upper) curve is from the equilibrium relation $\xi^{-1}=2\rho$, where  $\rho = \int_0^{2\pi}\frac{dk}{2\pi}n_k$ is the density of excitations. (b) Comparison between the numerically computed $\rho$ (upper dots) and the modified density $\rho_\xi$ (lower dots), defined in Eq.~(\ref{rho_xi}). The solid curves were obtained by assuming a parabolic dispersion and $n_k\simeq n_\pi$ within the occupied region. }
\end{figure}

Applying Eq.~(\ref{xi_Szego}) to the critical point, we first assume as in Fig.~\ref{fig:occupation_numbers}(a) that the minimum of the quasiparticle dispersion occurs at $k=\pi$. Furthermore, if $\mu_{\rm R}$ is inside the gap the critical point is at $\mu_{\rm L} = m_2$. Close to the critical point, the only occupied states are in a small range $k \in [\pi-\Delta k_{\rm L},\pi+\Delta k_{\rm L}]$ around the minimum of $\varepsilon_{k}$, thus we can approximate the quasiparticle dispersion as parabolic giving $\Delta k_L \propto \sqrt{\mu_{\rm L} - m_2}$. This dependence of $\Delta k_L$ is directly related to the critical exponent $\nu=1/2$. More precisely, $\Delta k_L$ close to the critical point is given by:
\begin{equation}
\Delta k_{\rm L} \simeq \sqrt{\frac{|1+h/J|}{\gamma^2-h/J-1}(\mu_{\rm L} - m_2)},
\label{eq: k_L expansion}
\end{equation}
and we can set $n_{k} \simeq n_{k=\pi}$ in the small integration interval of Eq.~(\ref{xi_Szego}), leading to:
\begin{equation}\label{xi_divergence_accurate}
\xi^{-1} \simeq  -\frac{\Delta k_{\rm L}}{\pi}\log\left[1-2n_{\pi}\right].
\end{equation}
This expression clearly shows how the divergence of $\xi$ is due to the shrinking of the region of nonzero occupation. We show in Fig.~\ref{fig: occupation number density} that this theory is accurate, by a direct comparison to the numerical results.

After having clarified the origin of the critical exponent $\nu=1/2$, it is interesting to compare the behavior of the open chain to the temperature dependence of the equilibrium system. In the latter case, an ordered phase is only allowed at zero temperature and the order disappears at any arbitrarily small temperature $T>0$. The sudden disappearance of the ordered state is related to the presence of thermal excitations, and is analogous to the vanishing of $\phi$ induced by the non-equilibrium chemical potentials, as soon as either $\mu_{\rm L}$ or $\mu_{\rm R}$ overcomes the gap. Furthermore, similarly to the non-equilibrium system, the correlation length diverges when $T \to 0$. As it turns out, in the low-temperature limit, $\xi$ can be related in a simple way to the density of excitations $\rho = \int_0^{2\pi}\frac{dk}{2\pi}n_k$:
\begin{equation}\label{eq: correlation length and rho} 
\xi^{-1}=2\rho. \qquad {\rm (low\mathrm{-}temperature)}
\end{equation}
This expression follows immediately from Eq.~(\ref{xi_Szego}), since $n_k=n_{-k} \ll 1$ when $T\to 0$,  and can also be understood by a simple argument in terms of a dilute gas of domain-wall excitations.\cite{Sachdev-2011} 

A naive application of Eq.~(\ref{eq: correlation length and rho}) to the non-equilibrium system is shown in Fig.~\ref{fig: occupation number density}. Although Eq.~(\ref{eq: correlation length and rho}) predicts the correct critical exponent $\nu=1/2$, there is a clear disagreement with the numerical results.  This failure of Eq.~(\ref{eq: correlation length and rho}) can be explained from the non-vanishing value of $n_k$ around the minimum of $\varepsilon_k$ (say, $k=\pi$): in both the low-temperature limit and the non-equilibrium system we have $\rho \to 0$ at the critical point, which results in a diverging correlation length. However, in the first case we have $n_k \to 0$ while for the non-equilibrium system $n_\pi$ remains finite, and the vanishing of $\rho $ is due to the shrinking of $\Delta k_L$. Instead of the density $\rho$, we can consider a `modified' density:
\begin{equation}\label{rho_xi}
\rho_\xi = -\frac{1}{4\pi}\int_{0}^{2\pi}\log\left[1-n_{k}-n_{-k}\right]dk,
\end{equation}
which follows naturally from Eq.~(\ref{xi_Szego}) by requiring that a relation similar to the equilibrium system at low temperature is satisfied, $\xi^{-1}=2 \rho_\xi$. It is easy to see that close the critical point we have $\rho_\xi \simeq  -(\log\left[1-2n_{\pi}\right]/2 n_\pi) \rho$, which differs from $\rho$ by a nontrivial multiplicative factor. An interesting exception, discussed more extensively in Appendix~\ref{app:analytical-approach-mixed-order}, occurs for $J = h/2$, when $n_\pi = 0$ and the relation between $\xi$ and $\rho$ is Eq.~(\ref{eq: correlation length and rho}), as in equilibrium. At $J = h/2$ the vanishing of $n_k$ also affects the value of the critical exponent, which is $\nu=5/2$ instead of 1/2. 

The above arguments can be adapted to other parameter regimes. In particular, the discussion is almost unchanged for $h/J < \min[0,\gamma^2-1]$, when the dispersion minimum is at $k=0$. Instead, the treatment of the sensitive phase with $|\gamma|<1$ is more delicate. Firstly, as shown in Fig.~\ref{fig:occupation_numbers}(b), the dispersion is characterized by two minima instead of one. More importantly, $n_k$ appears to be highly pathological when $N \to \infty$, when the occupation numbers undergo wild oscillations. Despite these differences, numerical evaluation of the critical exponent still gives $\nu=1/2$, thus we conjecture that a suitable average of $n_k$ is well-defined: 
\begin{equation}
\overline{n_k}=\lim_{\Delta k \to 0} \lim_{N\to \infty} \frac{1}{\Delta k}  \int_{k-\frac{\Delta k}{2}}^{k+\frac{\Delta k}{2}}  n_{k'} dk'.
\label{eq: nk expanded first order}
\end{equation} 
Then, the critical exponent would be determined through Eq.~(\ref{xi_divergence_accurate}) in a way analogous to the regular case, i.e., the critical exponent would correspond to the shrinking of the occupied regions around the minima, explaining the persistence of $\nu=1/2$ in this phase.

\subsection{z-correlations}

As summarized in Table~\ref{tb: table correlations}, $\mathbb{C}_{r,r'}^{zz}$ displays a power-law decay in several of the allowed phases. We focus here on the the non-sensitive region, where this behavior can be understood from the well-known power-law decay of density-density correlations of non-interacting fermions. In fact, through the fermionic mapping, $\mathbb{C}_{r,r'}^{zz} $ is equivalent to a density-density correlation function:
\begin{equation}
    \mathbb{C}_{r,r'}^{zz} = 4 \left( \langle \hat{c}_r^\dag \hat{c}_r \hat{c}_{r'}^\dag \hat{c}_{r'}\rangle -  \langle \hat{c}_r^\dag \hat{c}_r \rangle \langle \hat{c}_{r'}^\dag \hat{c}_{r'}\rangle \right).
\end{equation}
In the C phase, at least one of the reservoirs has its chemical potential within the range of quasiparticle energy spectrum. Then, the correlation function is expected to have a power-lay decay with oscillating character, similar to a simple 1D Fermi gas where it decays as $ |r-r'|^{-2}$ and oscillates with wavevector $2k_F$, with  $k_F$ the Fermi wavevector (see, e.g., Ref.~\onlinecite{Castin2007}). 

\begin{figure}[t]
\centering{}\includegraphics[width=0.9\columnwidth]{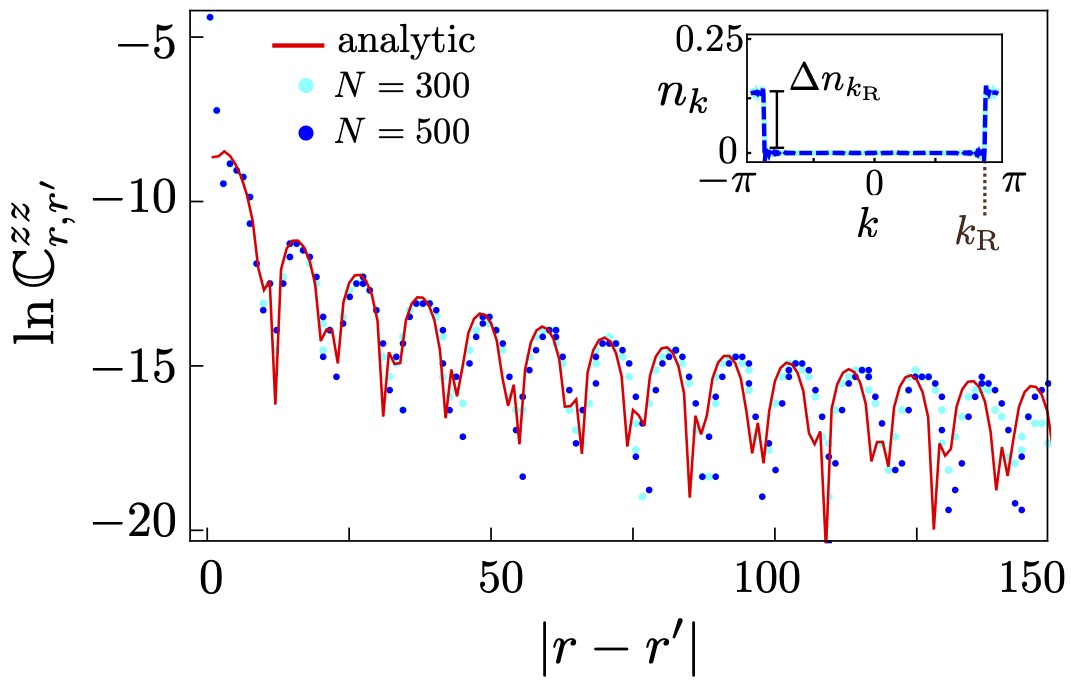}
\caption{\label{fig: analytic z-correlation} 
Comparison between numerical results for $\mathbb{C}_{r,r'}^{zz}$ (dots) and the analytic approximation Eq.~(\ref{Czz_single_FS}) (red curve). The values $\Delta n_{k_\text{R}} \simeq 0.1$ and $k_\text{R} \simeq 2.85$ were extracted numerically from $n_k$, shown in the inset. We used
$h=0.2$, $\gamma=1$, $\mu_{\text{L}} =0.5 < m_2$, and $\mu_{\text{R}} = 1.62 \gtrsim m_3$. 
}
\end{figure}

In our case, we express the correlation function through the Bogolubov excitations $\hat\gamma_k$ of the transnational invariant system (which is appropriate in the thermodynamic limit). The corresponding occupation numbers are defined in  Eq.~(\ref{eq:occupationo number definition}) and give:
\begin{align}\label{Czz_nk}
&\mathbb{C}_{r,0}^{zz} =\left\vert\int_{-\pi}^\pi \frac{dk}{2\pi}e^{ikr}
\left( n_{-k}+n_{k}-1\right) \sin 2\theta _{k}\right\vert ^{2}  \nonumber \\
& -\left\vert \int_{-\pi}^\pi \frac{dk}{2\pi} e^{ikr}\left[\left(
n_{k}+n_{-k}-1\right) \cos 2\theta _{k}+\left( n_{k}-n_{-k}\right) \right]
\right\vert ^{2}.
\end{align}
If for simplicity we assume $n_k \simeq n_{-k}$ (which is justified in the limit of vanishing hybridization energy $\Gamma$) the occupation numbers have discontinuities at $k=\pm k_{i}$, induced by the left and right reservoirs ($i=L,R$). When $k_{i} r\gg 1$, we can extract the leading contribution to Eq.~(\ref{Czz_nk})induced by the discontinuous jumps $\Delta n_{k_i}$ of $n_k$, defined by $\partial_k n_k = \sum_{i=L,R} \Delta n_{k_i}\left[\delta(k+k_i)-\delta(k-k_i)\right]$:
\begin{align}\label{Czz_two_FS}
\mathbb{C}_{r,0}^{zz} \simeq  \frac{4}{\pi ^{2}r^{2}} & \Bigg[ \bigg( \sum_{i=L,R}\Delta
n_{k_{i}}\sin 2\theta _{k_{i}}\cos k_{i}r\bigg) ^{2}  \nonumber \\
 - & \bigg(\sum_{i=L,R}\Delta n_{k_{i}}\cos 2\theta _{k_{i}} \sin k_{i}r\bigg) ^{2}%
\Bigg],
\end{align}
which simplifies to:
\begin{equation}\label{Czz_single_FS}
\mathbb{C}_{r,0}^{zz}  \simeq  \frac{4\Delta n_{k_{R}}^{2}}{\pi ^{2}r^{2}}\left( \cos ^{2}k_{R}r-\cos
^{2}2\theta _{k_{R}}\right),
\end{equation}
when there is a single Fermi surface (here, induced by $i=R$). The above expressions display the expected $1/r^2$ decay and oscillatory dependence. As shown in Fig.~\ref{fig: analytic z-correlation}, we find a good agreement between Eq.~(\ref{Czz_single_FS}) and the numerical results.

\begin{figure}[t]
\begin{centering}
\hfill{}\includegraphics[width=0.99\columnwidth]{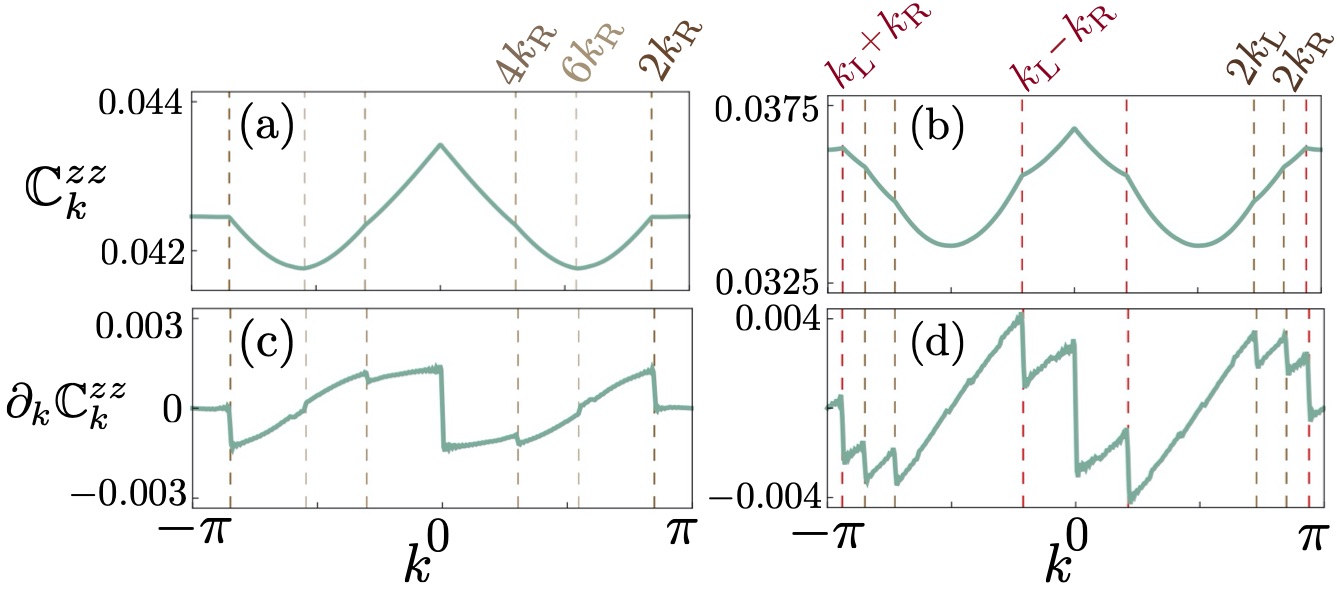}\hfill{}
\par\end{centering}
\centering{}\caption{\label{fig: z-correlation function momentum space}
Within the phase diagram of Fig. \ref{fig: non-Markovian phase diagram}(a), panels (a) and (b) show
the Fourier transform of $\mathbb{C}_{r,r'}^{zz}$ [see Eq.~(\ref{Czz_FT})] for points inside the C phase, 
namely (a) at $\{\mu_{\text{L}},\mu_{\text{R}}\}=\{-2.9,-2.1\}$ and (b) at $\{\mu_{\text{L}},\mu_{\text{R}}\}=\{-1.9,-2.1\}$.
(c) and (d) show their respective derivatives. In the left column we considered the case with only nonzero $k_{\text{R}}$,
while the right column considers nonzero $k_{\text{L}}$ and $k_{\text{R}}$.
We have computed these results for a system size of $N=500$ sites.}
\end{figure}

To better characterize the oscillatory dependence, we have also studied the Fourier transform: 
\begin{equation}\label{Czz_FT}
    \mathbb{C}_{k}^{zz}\equiv\frac{1}{\sqrt{N}}\sum_{r-r'}\text{e}^{-ik\left(r-r'\right)}\mathbb{C}_{r,r'}^{zz},
\end{equation}
which is shown in Fig.~\ref{fig: z-correlation function momentum space} for two representative cases. With a single Fermi surface (left panels) we find dominant non-analytic features at $k=0,2k_R$, in agreement with Eq.~(\ref{Czz_single_FS}). We also find smaller discontinuities in $ \partial_k \mathbb{C}_{k}^{zz}$ at higher harmonics, $k=4k_R,6k_R$, which are not captured by the leading-order approximation Eq.~(\ref{Czz_single_FS}). With two Fermi surfaces (right panels) we find the expected singularities at $k=0, 2k_{L,R}$. However, there are additional features at  $ \partial_k \mathbb{C}_{k}^{zz}$ at $k=k_L \pm k_R$, which are in agreement with Eq.~(\ref{Czz_two_FS}). As seen there, the correlation function is not simply a sum of $i=L,R$ contributions, but involves interference terms between the two Fermi surfaces.

Finally, we comment on the power-law dependence of $\mathbb{C}_{r,r'}^{zz}$ in the sensitive region. Away from the ordered phase we find a power-law decay $ |r-r'|^{-s}$ where, however, the exponent is generally different from $s=2$ (we often find $s<2$) and depends on system parameters. This behavior is most likely related to the singular nature of $n_k$, which from our numerical evidence is characterized by a complex pattern of closely spaced discontinuities (see, e.g., Fig.~\ref{fig:occupation_numbers}). Such discontinuities will contribute to the square parenthesis of Eq.~(\ref{Czz_two_FS}) in a way difficult to compute explicitly (the summation index $i$ should become a continuous parameter) and might be able to modify the exponent $s$. 
This interpretation is confirmed by the survival of the power-lay decay in the NC$^*$ and CS$^*$ regions, where the chemical potentials $\mu_{\rm L,R}$ do not cross the quasiparticle bands, thus an exponential decay might be expected.
Instead, Fig.~\ref{fig:occupation_numbers}(g) and (i) show that a discontinuous dependence of $n_k$  can be found in these regions as well, in agreement with the observed power-law dependence of  $\mathbb{C}_{r,r'}^{zz}$.
Finally, the simple discontinuities of Fig.~\ref{fig:occupation_numbers}(l) result in the regular value $s=2$.

\section{Entanglement entropy\label{sec:Entropy}}

In this section, we study the entropy of the steady-state within the different phases identified above. 
For a segment of $\ell$ sites in the middle of the chain, the entropy is given by
\begin{equation}
E_{\ell}=-\text{Tr}\left[\hat{\rho}_{\ell}\ln\left(\hat{\rho}_{\ell}\right)\right]=-\tr\left[\chi_{\ell}\ln\chi_{\ell}\right],
\end{equation}
where $\hat{\rho}_{\ell}$ is the reduced density matrix and $\chi_{\ell}$
is the single-particle correlation matrix restricted to the subsystem of
$\ell$ sites. While, for the fermionic system the second equality follows from the non-interacting nature of the problem, this expression was also shown to hold for the spin chain \cite{Vidal.03}. 
In the thermodynamic limit ($N\rightarrow\infty$)
the entropy of segment of a translational invariant system is expected to obey the general
scaling law\citep{Its.2009} 
\begin{equation}
E_{\ell}=l_{0}\ell+c_{0}\ln\left(\ell\right)+c_{1},\label{eq:entanglement entropy scaling law}
\end{equation}
where $l_{0}$, $c_0$ and $c_1$ are $\ell$-independent real constants.  
For the ground-state of gapped systems $l_{0}=c_{0}=0$ - following the so-called area law, while gapless fermions and spin chains show an universal logarithmic behavior with $c_{0}=1/3$. 
This result is a consequence of the violation of the area law in $1$+$1$ conformal
theories, in which case $c_{0}=c/3$, where $c$ is the central charge\citep{Vidal.03,Calabrese.04}.
For a nonequilibrium Fermi-gas, it was shown that both $l_{0}$ and
$c_{0}$ can be non-zero\citep{Eisler.14,Ribeiro.17}, and that $c_{0}$
depends on the system-reservoir coupling and is a non-analytic function
of the bias\citep{Ribeiro.17}.
The coefficient $c_{0}$ is most easily extracted
from the mutual information, ${\cal I}\left(A,B\right)\equiv E\left(\hat{\rho}_{A}\right)+E\left(\hat{\rho}_{B}\right)-E\left(\hat{\rho}_{A+B}\right)$,
of two adjacent segments $A$ and $B$ of size $\ell/2$, since 
\begin{equation}
{\cal I}_\ell\simeq c_{0}\ln\left(\ell\right)+c_{2}.\label{eq: mutual information scaling law}
\end{equation}

For the transverse-field Ising model, in Fig. \ref{fig: non-Markovian phase diagram}(a),
all phases, except O, have been shown to have extensive entropy
(i.e. $l_{0}\neq 0$) \citep{Puel-Chesi-Kirchner-Ribeiro-2019}. This is due to the presence of a finite fraction of excitations, which are absent in the ordered phase. 
In addition, it was found that $c_{0}\neq 0$ in the C phase, due to the presence of discontinuities in $n_{k}$. 

In Fig.\ref{fig: entropy scaling law} we show the generalization of the previous results to the $XY$-chain and including the sensitive region of the phase diagram. 
Fig.\ref{fig: entropy scaling law} (a) shows that $l_{0}$ for both normal and sensitive regions. In both cases $l_{0}$ follows the expected value \citep{Ribeiro.17}:
\begin{equation}
l_{0} =   \int \frac{dk}{2\pi} - \left[ n_k \ln n_k + (1-n_k) \ln (1-n_k)\right].\label{eq:entropy}
\end{equation}
On the other hand, ${\cal I}_\ell$ does differ qualitatively in the normal and sensitive regions. 
As aforementioned, logarithmic corrections come from discontinuities in $n_{k}$. 
If $n_{k}$ has no discontinuities such as in the O and saturated normal phases, $c_{0}$ vanishes. 
Fig.\ref{fig: entropy scaling law} (b) depicts $c_{0}$ in the normal region.
The right inset shows that the leading term of ${\cal I}_\ell$ in Eq.(\ref{eq: mutual information scaling law}) is indeed logarithmic in the large $\ell$ limit. 
In principle, for conducting non-saturated phases in the normal region, $c_{0}$ can be computed using the Fisher-Hartwing conjecture \cite{Its.2009}. 

The presence of noise in $n_k$ within the sensitive region changes the previous picture. 
The right inset of Fig.\ref{fig: entropy scaling law} (b) shows that ${\cal I}_\ell$ is no longer of the form given in Eq.(\ref{eq: mutual information scaling law}). 
It is tempting to interpret this result as a collection of discontinuities of $n_k$ whose number increases with system size. However, due to the noisy form of $n_k$ it is difficult to give a precise meaning to this picture. Moreover, with the system sizes we could attain, it was not possible to determine the functional for of this correction wit $\ell$.   
 Note that, a supra-logarithmic contribution in ${\cal I}_\ell$ is always present in the sensitive region even for the CS$^*$ and NC$^*$ phases.

\begin{center}
\begin{figure}[t]
\centering{}\includegraphics[width=0.98\columnwidth]{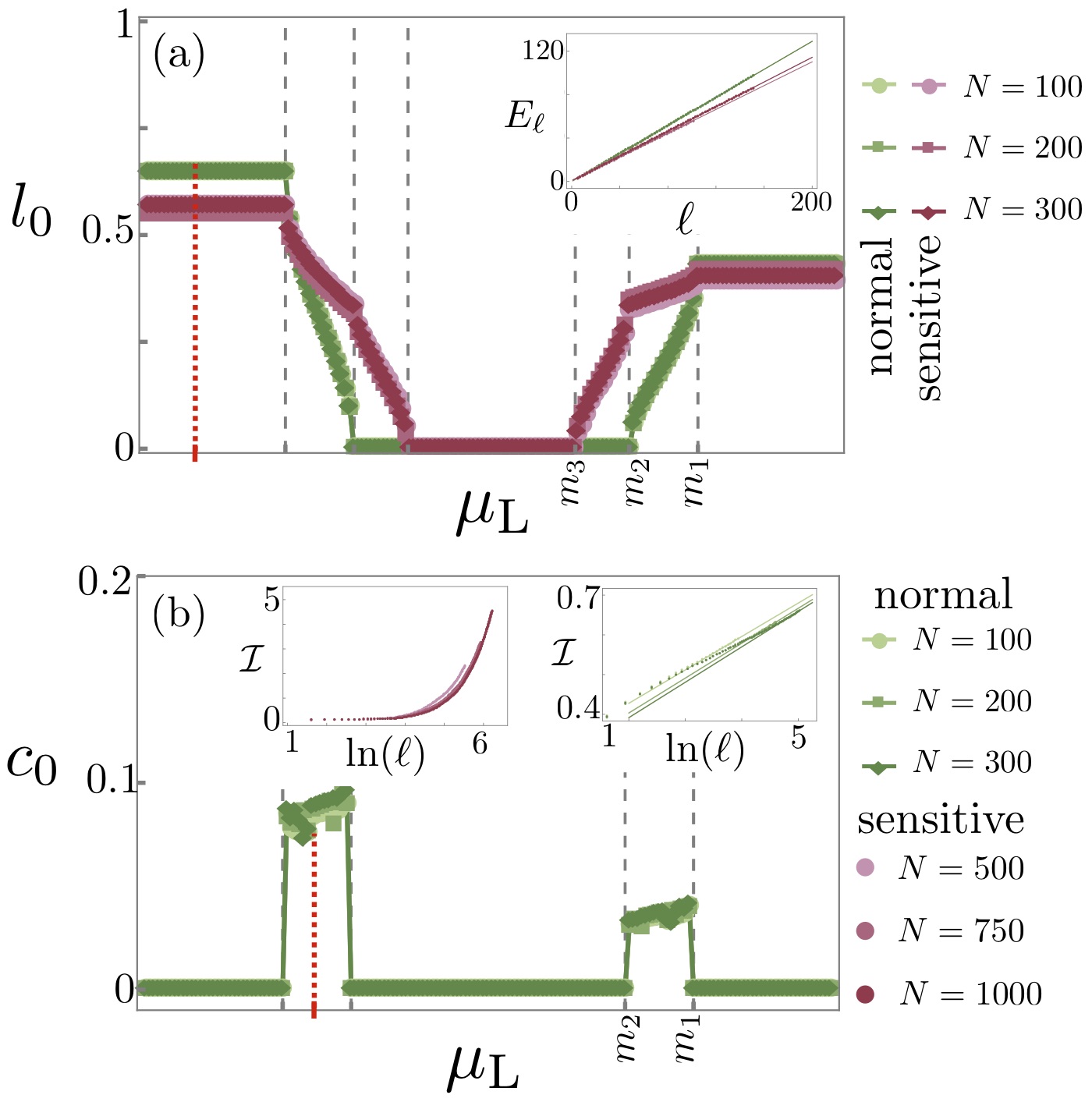}\caption{\label{fig: entropy scaling law} (a) Linear coefficient $l_{0}$,
obtained from the entropy scaling law, for a range of $\mu_{\text{L}}$
across the lines depicted in the phase diagrams of Fig. \ref{fig: non-Markovian phase diagram}.
The inset illustrates the typical fitting $E_{\ell}\times\ell$, where
the value of $\mu_{\text{L}}$ used is denoted by the dotted-red
line. (b) Shows similar results for the logarithmic coefficient $c_{0}$,
obtained from Eq.(\ref{eq: mutual information scaling law}), for
the same range of $\mu_{\text{L}}$. The inset on the right illustrates
the typical fitting ${\cal I}\times\ln\left(\ell\right)$, where the
value of $\mu_{\text{L}}$ is denoted by the dotted-red line. In
the sensitive region the mutual information is super-logarithmic. }
\end{figure}
\par\end{center}

\section{Discussions\label{sec:Discussions}}

We have studied the steady-state of a transverse field $XY$-spin chain at zero temperature in a non-equilibrium setting by coupling the ends of the chain to reservoirs which can be held at different magnetic potentials. 
Our approach is based on a Jordan-Wigner mapping and  a Keldysh Green's function treatment of the resulting non-equilibrium interaction-free fermionic system. 

This allows us to study steady-states of the model as a function of the magnetic potentials including the equilibrium and Markovian limits.
For magnetic potentials whose magnitudes remain smaller than the spectral gap,
the equilibrium-ordered state persists and the correlation function of the order parameter displays long-range order. Away from this phase, the order parameter correlations decay exponentially. 
As $\mu_\text{L}$ or $\mu_\text{R}$ reaches the spectral gap, the transition from equilibrium to a current-carrying state occurs through a mixed-ordered transition, where the order parameter vanishes discontinuously while the correlation length diverges. 
This out-of-equilibrium phenomena was first observed in the Ref.  [\onlinecite{Puel-Chesi-Kirchner-Ribeiro-2019}]. Our present results establish that 
this behavior is generic to to all order/disorder phase transitions of the $XY$ chain. 

For large $\abs{\mu_\text{L}}$ and $\abs{\mu_\text{R}}$ we recover the Markovian limit. 
We identify the two qualitatively different behaviors previously reported using a Lindblad Master equation approach \cite{Prosen2008,Banchi2013}. 
We refer to these as  (i) sensitive region, featuring algebraic decaying correlations in the transverse (i.e. $z$) direction, and (ii) normal region, where these correlations decay exponentially. 

In addition to the equilibrium and Markovian phases, we identify new current-carrying phases.  Their properties can be easily understood by studying the quasi-particle excitation number $n_k$.
By analysing this quantity, we were able to compute the critical exponent of the diverging correlation length at the transition, confirming analytically the results of Ref. [\onlinecite{Puel-Chesi-Kirchner-Ribeiro-2019}].  
The behavior of the transverse correlation function in the normal region is explained in terms of a physical effect which is similar to Friedel oscillations in metals, here observed in a non-equilibrium setting.  

For steady-state phases within the sensitive region $n_k$ is noisy, for $k$ belonging to the intervals of momentum where the dispersion relation allows four propagating modes. This noise cannot be interpreted as  a finite size feature since $n_k$, within these regions, does not converge to a thermodynamic limit. Since the transverse correlation function is related to the Fourier transform of $n_k$, the pathologies of this function explain the non-exponential decay of transverse correlations in the sensitive region. 
 
We have also analyzed the behavior of the entropy of a segment of the steady-state with its length. 
As expected for a mixed state, the steady-state entropy is extensive and follows its predicted semi-classical value. 
In the normal region, whenever the chemical potential lies within one of the bands, there is a logarithmic component that is  reminiscent of the area-law violation occurring  in equilibrium gapless states. As reported for other non-equilibrium setups \cite{Ribeiro.17}, the logarithmic coefficient depends on the discontinuities of $n_k$. 
In the sensitive region, corrections to the extensive contribution turn out to be super-logarithmic.  

The analysis presented here generalizes our earlier findings on the transverse field Ising case to the anisotropic XY chain and, thus, extends the class of spin chains which display far-from-equilibrium critical behavior, reflected in  a divergent correlation length, that is absent in  equilibrium. Thus, the present work suggests that these findings may reflect  common features of a wide class of quantum statistical models.  A common feature of the models discussed in the present context is their equivalence to interaction-free fermions under a Jordan-Wigner transformation. This naturally poses the question if there exists a finite region in model space around these XY chains where similar non-equilibrium behavior ensures or if their non-thermal behavior is singular.  The Jordan-Wigner transformation is limited to one-dimensional systems. Yet, it would be desirable to understand how the non-equilibrium phases that we have identified generalize in higher dimensions.

In equilibrium, interaction-free or quadratic models are commonly associated with  fixed point behavior within a field-theoretic description of criticality. This enables one to  categorize  a wide class of systems into universality classes, with respect to the fixed points. Away from equilibrium, such a categorization is not available. 
Our results thus offer a vantage point for the construction of  a wider class of models that share the same out-of-equilibrium behavior. A better understanding of the universality of far-from-equilibrium critical behavior should prove beneficial for the construction of a field theoretic description of quantum critical matter far from equilibrium.

\begin{acknowledgments}
We gratefully acknowledge helpful discussions with T. Prosen, V.R.
Vieira, and R. Fazio. P. Ribeiro acknowledges support by FCT through Grant No. UID/CTM/04540/2019.
S. Kirchner acknowledges support by the National Science Foundation
of China, grant No. 11774307 and the National Key R\&D Program of
the MOST of China, Grant No. 2016YFA0300202. S. Chesi acknowledges
support from NSFC (Grants No. 11574025 and No. 11750110428).
\end{acknowledgments}


\appendix

\section{Method detailed\label{app:Method-detailed}}

The single-particle density matrix in Eq. (\ref{eq: single-particle density matrix})
is explicitly given by

\begin{multline}
\bs{\chi}=\frac{1}{2}+\sum_{l=\t L,\t R}\sum_{\alpha\beta}\ket{\alpha}\bra{\beta}\times\\
\bra{\tilde{\alpha}}\left[\bs{\gamma}_{l}I_{l}\left(\lambda_{\alpha},\lambda_{\beta}^{*}\right)-\hat{\bs{\gamma}}_{l}I_{l}\left(-\lambda_{\alpha},-\lambda_{\beta}^{*}\right)\right]\ket{\tilde{\beta}}\label{eq:chi}
\end{multline}
 where $I_{l}\left(z,z'\right)=-\frac{1}{\pi}\frac{g\left(z-2m_{l}\right)-g\left(z'-2m_{l}\right)}{z-z'}$
with $g\left(z\right)=\ln\left(-i\sgn\left[\im\left(z\right)\right]z\right)$,
and the matrices $\bs{\gamma}_{l}$ are defined in Eq. (\ref{eq: K matrix}).
Here we assumed that $\bs K$ in Eq. (\ref{eq: K matrix}) is diagonalizable,
having right and left eigenvectors $\ket{\alpha}$ and $\bra{\tilde{\alpha}}$
with associated eigenvalues $\lambda_{\alpha}$.

The energy drained to the left reservoir is $\mathcal{J}_{e}=-i\left\langle \left[H,H_{\t L}\right]\right\rangle $,
which equals the steady-state energy current in any cross section
along the chain and thus can be obtained as a function of $\bs{\chi}$.
Explicitly the energy flow can be obtained as $\mathcal{J}_{e}=-\frac{1}{2}\,\text{Tr}\left[\boldsymbol{J}_{r}\boldsymbol{\chi}\right]$
$\forall\,r$, Eq. (\ref{eq: current of energy}), with
\begin{multline}
\boldsymbol{J}_{r}=-2ihJ\left[(1+{\bs S})\ket{r-1}\bra{r}(1+{\bs S})-\text{H.c.}\right]\,.
\end{multline}
 The linear and non-linear thermal conductivities, as well as other
thermoelectric properties of the chain, are determined by $\mathcal{J}_{e}$.

\subsection{Two-points correlation} 

Let us further analyse the two-points correlation function in Eq.
(\ref{eq: correlation definition}), which can also be found in terms
of $\bs{\chi}$. To this end we have extended the equilibrium expressions\citep{Lieb1961}
to general non-equilibrium conditions: 
\begin{eqnarray}
\mathbb{C}_{r,r'}^{xx} & = & \det\left[i\left(2\bs{\chi}_{\left[r,r'\right]}-1\right)\right]^{\frac{1}{2}}\label{eq:C_xx}
\end{eqnarray}
for $r>r'+1$, where $\bs{\chi}_{\left[r,r'\right]}$ is a $2\left(r-r'\right)$
matrix obtained as the restriction of $\bs{\chi}$ to the subspace
in which $\mathbb{P}_{rr'}^{T}=\sum_{u=r'+1}^{r-1}\left(\ket u\bra u+\ket{\hat{u}}\bra{\hat{u}}\right)+\ket{r_{+}}\bra{r_{+}}+\ket{r'_{-}}\bra{r'_{-}}$,
with $\ket{r_{\pm}}=\left(\ket r\pm\ket{\hat{r}}\right)/\sqrt{2}$,
acts as the identity,
and $\ket r$ and $\ket{\hat{r}}\equiv\bs S\ket r$ are single-particle  and hole states.

\section{Excitation number \label{app: excitation number}}

\begin{figure}[t]
\centering{}\includegraphics[width=0.75\columnwidth]{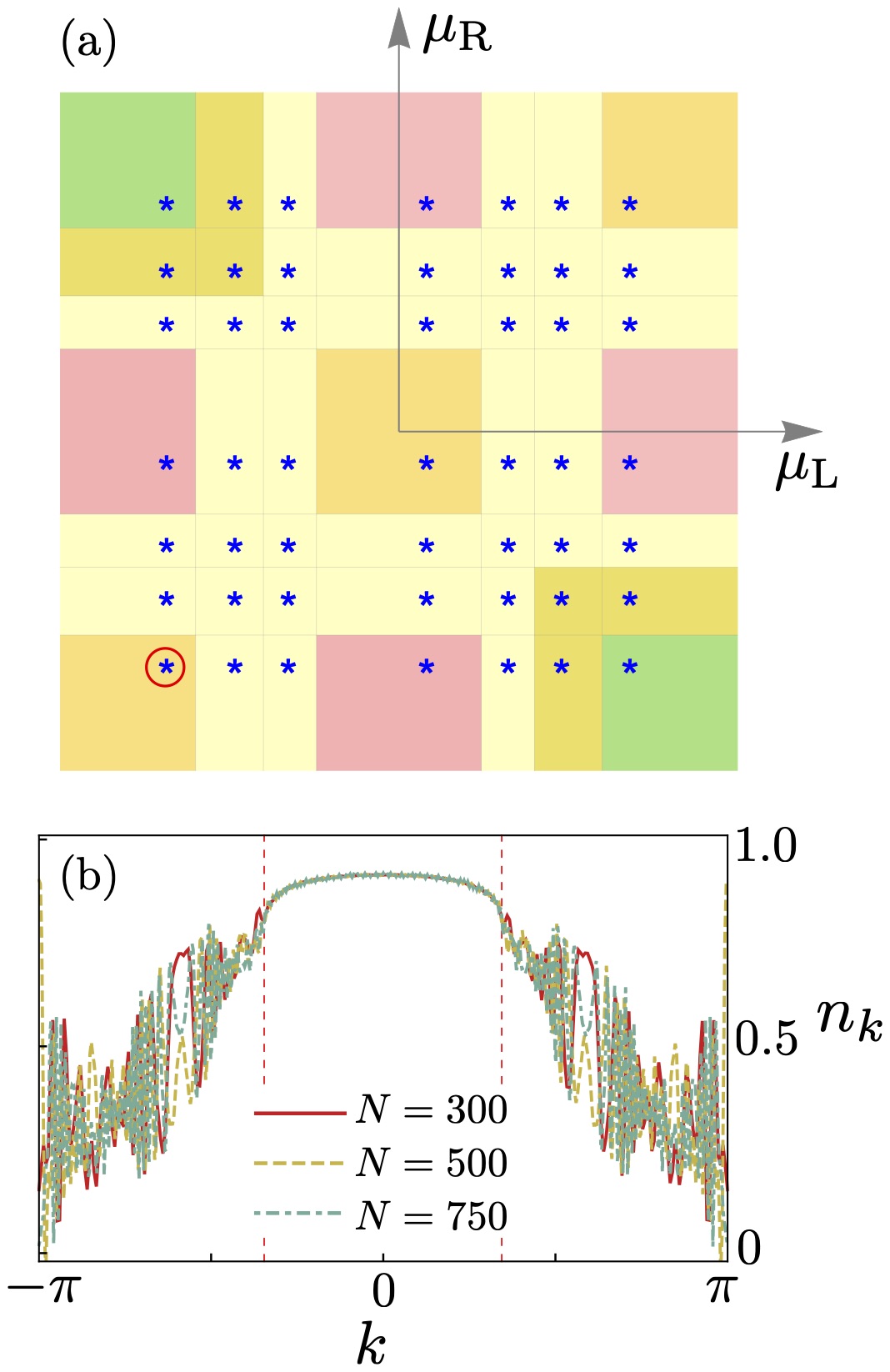}\caption{\label{fig: points in the phase diagram} (a) repeats the phase diagram
of Fig. \ref{fig: non-Markovian phase diagram}(b). The blue stars
mark the parameters used in Fig. \ref{fig: excitation number appendix}.
Panel (b) illustrates the persistent noise in the occupation number,
computed for different system sizes, with set of parameters indicated
by the circled blue star on the phase diagram.}
\end{figure}

\begin{figure*}
\centering{}\includegraphics[width=0.95\textwidth]{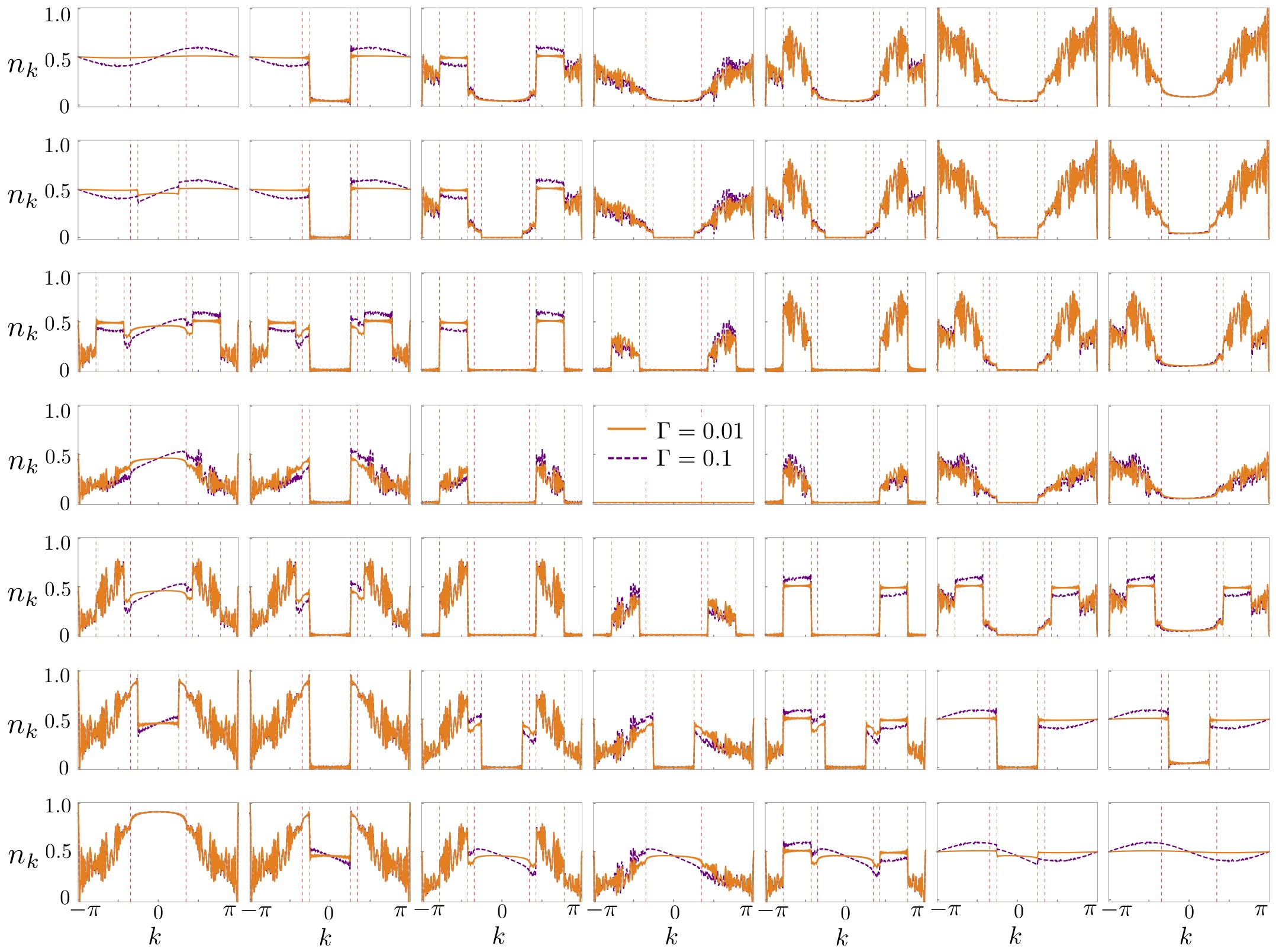}\caption{\label{fig: excitation number appendix} Excitation number $n_{k}$
in the sensitive region for different phases. Each plot represents
a different set of parameters $\left\{ \mu_{\text{L}},\mu_{\text{R}}\right\} $
marked on the phase diagram in Fig. \ref{fig: points in the phase diagram}(a)
disposed in the same order. We have computed it for a system size
of $N=500$ sites. The vertical dashed lines represent either the
momentum $k_{m_{2}}$ or $k_{\text{L}}$, see Figs. \ref{fig: band spectrum and rapidities}(c)
and \ref{fig:occupation_numbers}(a) and \ref{fig:occupation_numbers}(b).}
\end{figure*}

The intriguing results of the occupation number, computed in section
\ref{sec:Phase-diagram}D, deserve a more detail analysis as follows.
As discussed in the manuscript, the occupation number shows a noise
behavior in some of the phases, thus, in order to give the reader
a complete view of the possible cases, in Fig. \ref{fig: excitation number appendix}
we show the occupation number for one representative point in each
phase, as marked in Fig. \ref{fig: points in the phase diagram}(a).
Here we have set the system in the sensitive region with transverse
magnetic field $h/J=0.2$ and anisotropy $\gamma=0.5$. 
The marks are at the points $\mu_\text{L},\mu_\text{R} = \{\pm2.7,\pm1.9,\pm1.3,+0.3 \text{ or} -0.3\}$.

As discussed in the manuscript, here we clearly see that the noise
is present on the phases around the axis $\mu_{\text{L}}=\mu_{\text{R}}$,
while it vanishes around the axis $\mu_{\text{L}}=-\mu_{\text{R}}$.
The noise only appears for $\left|k\right|>k_{m_{2}}$, see Fig. \ref{fig: band spectrum and rapidities}(c).
In addition, it is persistent even in the thermodynamic limit, as
shows Fig. \ref{fig: points in the phase diagram}(b).

\pagebreak{}

\section{Correlation length at $h/J=0.5$\label{app:analytical-approach-mixed-order}}

\begin{figure}
\includegraphics[width=0.98\columnwidth]{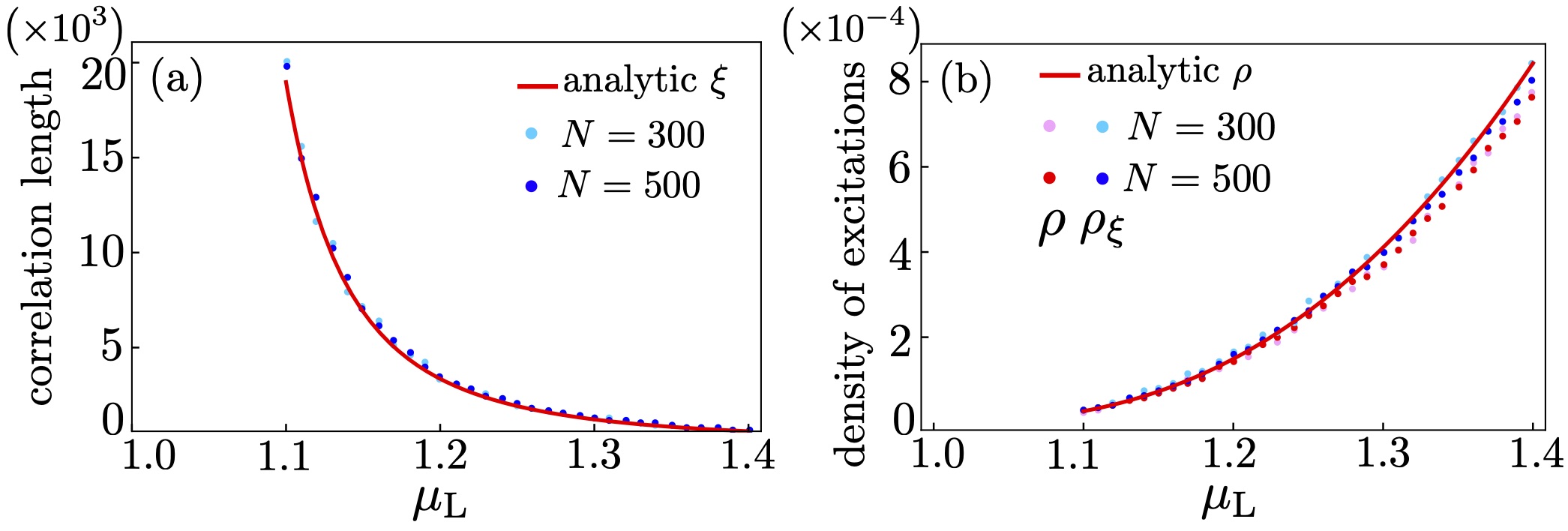}
\caption{\label{fig: occupation number density special case h 0.5} (a) Correlation
length of the Ising model ($\gamma=1$) for the special case $h/J=0.5$
and approaching the critical point $m_{2}=1$ from the disordered
phase (C), with $\mu_\text{R} = 0$. The dots and continuous line compare the numerical
result, Eq. (\ref{eq:correlation and correlation length}), to the
analytic result, Eq. (\ref{xi_1/2}),
respectively. (b) Density of excitations  $\rho$ and $\rho_\xi$, given by Eq.~(\ref{rho_xi}). The continuous
line shows the analytic result for $\rho$, computed from Eq.~(\ref{xi_1/2}) as $\rho=\xi^{-1}/2$.
In the analytic formulas we have used the educated guess $c=\pi/5!$.}
\end{figure}

When $h/J=0.5$, it was observed that $n_{k=\pi}=0$,\cite{Puel-Chesi-Kirchner-Ribeiro-2019} thus the derivation of Eq.~(\ref{xi_divergence_accurate}) should be modified. Interestingly, around $k=\pi$ we find that $n_k$ has a leading term of the form:
\begin{equation}
n_{k}\simeq c (k-\pi)^{4},
\end{equation}
where the vanishing of the quadratic term and the value of $c$ could only be obtained numerically. In the limit of a small $\Delta k_{\rm L}$, Eq.~(\ref{xi_Szego}) yields:
\begin{equation}\label{xi_1/2}
\xi^{-1} \simeq  \frac{c}{\pi}\int_{-\Delta k_{\rm L}}^{\Delta k_{\rm L}} x^4 dx = \frac{2c}{5\pi} \Delta k_{\rm L}^5.
\end{equation}
Finally, from the expression of $\Delta k_{\rm L}$ in Eq.~(\ref{eq: k_L expansion}) we immediately see that the critical exponent is $\nu=5/2$. It is also worth mentioning that, since $n_{k}$ becomes vanishingly small approaching the critical point, Eq.~(\ref{rho_xi}) coincides with the regular density and the relation $\xi^{-1}=2\rho$ is still valid. Figure~\ref{fig: occupation number density special case h 0.5} shows
the comparison of these results with the numerical calculations.

\cleardoublepage{}


\bibliographystyle{apsrev4-1}
\bibliography{refs}

\begin{thebibliography}{52}%
\makeatletter
\providecommand \@ifxundefined [1]{%
 \@ifx{#1\undefined}
}%
\providecommand \@ifnum [1]{%
 \ifnum #1\expandafter \@firstoftwo
 \else \expandafter \@secondoftwo
 \fi
}%
\providecommand \@ifx [1]{%
 \ifx #1\expandafter \@firstoftwo
 \else \expandafter \@secondoftwo
 \fi
}%
\providecommand \natexlab [1]{#1}%
\providecommand \enquote  [1]{``#1''}%
\providecommand \bibnamefont  [1]{#1}%
\providecommand \bibfnamefont [1]{#1}%
\providecommand \citenamefont [1]{#1}%
\providecommand \href@noop [0]{\@secondoftwo}%
\providecommand \href [0]{\begingroup \@sanitize@url \@href}%
\providecommand \@href[1]{\@@startlink{#1}\@@href}%
\providecommand \@@href[1]{\endgroup#1\@@endlink}%
\providecommand \@sanitize@url [0]{\catcode `\\12\catcode `\$12\catcode
  `\&12\catcode `\#12\catcode `\^12\catcode `\_12\catcode `\%12\relax}%
\providecommand \@@startlink[1]{}%
\providecommand \@@endlink[0]{}%
\providecommand \url  [0]{\begingroup\@sanitize@url \@url }%
\providecommand \@url [1]{\endgroup\@href {#1}{\urlprefix }}%
\providecommand \urlprefix  [0]{URL }%
\providecommand \Eprint [0]{\href }%
\providecommand \doibase [0]{http://dx.doi.org/}%
\providecommand \selectlanguage [0]{\@gobble}%
\providecommand \bibinfo  [0]{\@secondoftwo}%
\providecommand \bibfield  [0]{\@secondoftwo}%
\providecommand \translation [1]{[#1]}%
\providecommand \BibitemOpen [0]{}%
\providecommand \bibitemStop [0]{}%
\providecommand \bibitemNoStop [0]{.\EOS\space}%
\providecommand \EOS [0]{\spacefactor3000\relax}%
\providecommand \BibitemShut  [1]{\csname bibitem#1\endcsname}%
\let\auto@bib@innerbib\@empty
\bibitem [{\citenamefont {Pothier}\ \emph {et~al.}(1997)\citenamefont
  {Pothier}, \citenamefont {Gu{\'{e}}ron}, \citenamefont {Birge}, \citenamefont
  {Esteve},\ and\ \citenamefont {Devoret}}]{Pothier1997}%
  \BibitemOpen
  \bibfield  {author} {\bibinfo {author} {\bibfnamefont {H.}~\bibnamefont
  {Pothier}}, \bibinfo {author} {\bibfnamefont {S.}~\bibnamefont
  {Gu{\'{e}}ron}}, \bibinfo {author} {\bibfnamefont {N.~O.}\ \bibnamefont
  {Birge}}, \bibinfo {author} {\bibfnamefont {D.}~\bibnamefont {Esteve}}, \
  and\ \bibinfo {author} {\bibfnamefont {M.~H.}\ \bibnamefont {Devoret}},\
  }\href {\doibase 10.1103/PhysRevLett.79.3490} {\bibfield  {journal} {\bibinfo
   {journal} {Phys. Rev. Lett.}\ }\textbf {\bibinfo {volume} {79}},\ \bibinfo
  {pages} {3490} (\bibinfo {year} {1997})}\BibitemShut {NoStop}%
\bibitem [{\citenamefont {Anthore}\ \emph {et~al.}(2003)\citenamefont
  {Anthore}, \citenamefont {Pierre}, \citenamefont {Pothier},\ and\
  \citenamefont {Esteve}}]{Anthore.03}%
  \BibitemOpen
  \bibfield  {author} {\bibinfo {author} {\bibfnamefont {A.}~\bibnamefont
  {Anthore}}, \bibinfo {author} {\bibfnamefont {F.}~\bibnamefont {Pierre}},
  \bibinfo {author} {\bibfnamefont {H.}~\bibnamefont {Pothier}}, \ and\
  \bibinfo {author} {\bibfnamefont {D.}~\bibnamefont {Esteve}},\ }\href
  {\doibase 10.1103/PhysRevLett.90.076806} {\bibfield  {journal} {\bibinfo
  {journal} {Phys. Rev. Lett.}\ }\textbf {\bibinfo {volume} {90}},\ \bibinfo
  {pages} {076806} (\bibinfo {year} {2003})}\BibitemShut {NoStop}%
\bibitem [{\citenamefont {Chen}\ \emph {et~al.}(2009)\citenamefont {Chen},
  \citenamefont {Dirks}, \citenamefont {Al-Zoubi}, \citenamefont {Birge},\ and\
  \citenamefont {Mason}}]{Chen.09}%
  \BibitemOpen
  \bibfield  {author} {\bibinfo {author} {\bibfnamefont {Y.-F.}\ \bibnamefont
  {Chen}}, \bibinfo {author} {\bibfnamefont {T.}~\bibnamefont {Dirks}},
  \bibinfo {author} {\bibfnamefont {G.}~\bibnamefont {Al-Zoubi}}, \bibinfo
  {author} {\bibfnamefont {N.~O.}\ \bibnamefont {Birge}}, \ and\ \bibinfo
  {author} {\bibfnamefont {N.}~\bibnamefont {Mason}},\ }\href {\doibase
  10.1103/PhysRevLett.102.036804} {\bibfield  {journal} {\bibinfo  {journal}
  {Phys. Rev. Lett.}\ }\textbf {\bibinfo {volume} {102}},\ \bibinfo {pages}
  {036804} (\bibinfo {year} {2009})}\BibitemShut {NoStop}%
\bibitem [{\citenamefont {Brantut}\ \emph {et~al.}(2013)\citenamefont
  {Brantut}, \citenamefont {Grenier}, \citenamefont {Meineke}, \citenamefont
  {Stadler}, \citenamefont {Krinner}, \citenamefont {Kollath}, \citenamefont
  {Esslinger},\ and\ \citenamefont {Georges}}]{Brantut2013}%
  \BibitemOpen
  \bibfield  {author} {\bibinfo {author} {\bibfnamefont {J.-P.}\ \bibnamefont
  {Brantut}}, \bibinfo {author} {\bibfnamefont {C.}~\bibnamefont {Grenier}},
  \bibinfo {author} {\bibfnamefont {J.}~\bibnamefont {Meineke}}, \bibinfo
  {author} {\bibfnamefont {D.}~\bibnamefont {Stadler}}, \bibinfo {author}
  {\bibfnamefont {S.}~\bibnamefont {Krinner}}, \bibinfo {author} {\bibfnamefont
  {C.}~\bibnamefont {Kollath}}, \bibinfo {author} {\bibfnamefont
  {T.}~\bibnamefont {Esslinger}}, \ and\ \bibinfo {author} {\bibfnamefont
  {A.}~\bibnamefont {Georges}},\ }\href {\doibase 10.1126/science.1242308}
  {\bibfield  {journal} {\bibinfo  {journal} {Science}\ }\textbf {\bibinfo
  {volume} {342}},\ \bibinfo {pages} {713} (\bibinfo {year}
  {2013})}\BibitemShut {NoStop}%
\bibitem [{\citenamefont {B\"uttiker}\ and\ \citenamefont
  {Landauer}(1982)}]{PhysRevLett.49.1739}%
  \BibitemOpen
  \bibfield  {author} {\bibinfo {author} {\bibfnamefont {M.}~\bibnamefont
  {B\"uttiker}}\ and\ \bibinfo {author} {\bibfnamefont {R.}~\bibnamefont
  {Landauer}},\ }\href {\doibase 10.1103/PhysRevLett.49.1739} {\bibfield
  {journal} {\bibinfo  {journal} {Phys. Rev. Lett.}\ }\textbf {\bibinfo
  {volume} {49}},\ \bibinfo {pages} {1739} (\bibinfo {year}
  {1982})}\BibitemShut {NoStop}%
\bibitem [{\citenamefont {Stefanucci}\ and\ \citenamefont {van
  Leeuwen}(2013)}]{stefanucci_vanleeuwen_2013}%
  \BibitemOpen
  \bibfield  {author} {\bibinfo {author} {\bibfnamefont {G.}~\bibnamefont
  {Stefanucci}}\ and\ \bibinfo {author} {\bibfnamefont {R.}~\bibnamefont {van
  Leeuwen}},\ }\href {\doibase 10.1017/CBO9781139023979} {\emph {\bibinfo
  {title} {Nonequilibrium Many-Body Theory of Quantum Systems: A Modern
  Introduction}}}\ (\bibinfo  {publisher} {Cambridge University Press},\
  \bibinfo {year} {2013})\BibitemShut {NoStop}%
\bibitem [{\citenamefont {Datta}\ \emph {et~al.}(1997)\citenamefont {Datta},
  \citenamefont {Ahmad},\ and\ \citenamefont {Pepper}}]{datta1997electronic}%
  \BibitemOpen
  \bibfield  {author} {\bibinfo {author} {\bibfnamefont {S.}~\bibnamefont
  {Datta}}, \bibinfo {author} {\bibfnamefont {H.}~\bibnamefont {Ahmad}}, \ and\
  \bibinfo {author} {\bibfnamefont {M.}~\bibnamefont {Pepper}},\ }\href
  {https://books.google.com.br/books?id=28BC-ofEhvUC} {\emph {\bibinfo {title}
  {Electronic Transport in Mesoscopic Systems}}},\ Cambridge Studies in
  Semiconductor Physi\ (\bibinfo  {publisher} {Cambridge University Press},\
  \bibinfo {year} {1997})\BibitemShut {NoStop}%
\bibitem [{\citenamefont {Imry}(2002)}]{imry2002introduction}%
  \BibitemOpen
  \bibfield  {author} {\bibinfo {author} {\bibfnamefont {Y.}~\bibnamefont
  {Imry}},\ }\href {https://books.google.com.br/books?id=ZyjW37iGhaQC} {\emph
  {\bibinfo {title} {Introduction to Mesoscopic Physics}}},\ Mesoscopic physics
  and nanotechnology\ (\bibinfo  {publisher} {Oxford University Press},\
  \bibinfo {year} {2002})\BibitemShut {NoStop}%
\bibitem [{\citenamefont {Aoki}\ \emph {et~al.}(2014)\citenamefont {Aoki},
  \citenamefont {Tsuji}, \citenamefont {Eckstein}, \citenamefont {Kollar},
  \citenamefont {Oka},\ and\ \citenamefont {Werner}}]{RevModPhys.86.779}%
  \BibitemOpen
  \bibfield  {author} {\bibinfo {author} {\bibfnamefont {H.}~\bibnamefont
  {Aoki}}, \bibinfo {author} {\bibfnamefont {N.}~\bibnamefont {Tsuji}},
  \bibinfo {author} {\bibfnamefont {M.}~\bibnamefont {Eckstein}}, \bibinfo
  {author} {\bibfnamefont {M.}~\bibnamefont {Kollar}}, \bibinfo {author}
  {\bibfnamefont {T.}~\bibnamefont {Oka}}, \ and\ \bibinfo {author}
  {\bibfnamefont {P.}~\bibnamefont {Werner}},\ }\href {\doibase
  10.1103/RevModPhys.86.779} {\bibfield  {journal} {\bibinfo  {journal} {Rev.
  Mod. Phys.}\ }\textbf {\bibinfo {volume} {86}},\ \bibinfo {pages} {779}
  (\bibinfo {year} {2014})}\BibitemShut {NoStop}%
\bibitem [{\citenamefont {Gutman}\ \emph {et~al.}(2008)\citenamefont {Gutman},
  \citenamefont {Gefen},\ and\ \citenamefont {Mirlin}}]{Gutman.2008}%
  \BibitemOpen
  \bibfield  {author} {\bibinfo {author} {\bibfnamefont {D.~B.}\ \bibnamefont
  {Gutman}}, \bibinfo {author} {\bibfnamefont {Y.}~\bibnamefont {Gefen}}, \
  and\ \bibinfo {author} {\bibfnamefont {A.~D.}\ \bibnamefont {Mirlin}},\
  }\href {\doibase 10.1103/PhysRevLett.101.126802} {\bibfield  {journal}
  {\bibinfo  {journal} {Phys. Rev. Lett.}\ }\textbf {\bibinfo {volume} {101}},\
  \bibinfo {pages} {126802} (\bibinfo {year} {2008})}\BibitemShut {NoStop}%
\bibitem [{\citenamefont {Gutman}\ \emph {et~al.}(2009)\citenamefont {Gutman},
  \citenamefont {Gefen},\ and\ \citenamefont {Mirlin}}]{Gutman.2009}%
  \BibitemOpen
  \bibfield  {author} {\bibinfo {author} {\bibfnamefont {D.~B.}\ \bibnamefont
  {Gutman}}, \bibinfo {author} {\bibfnamefont {Y.}~\bibnamefont {Gefen}}, \
  and\ \bibinfo {author} {\bibfnamefont {A.~D.}\ \bibnamefont {Mirlin}},\
  }\href {\doibase 10.1103/PhysRevB.80.045106} {\bibfield  {journal} {\bibinfo
  {journal} {Phys. Rev. B}\ }\textbf {\bibinfo {volume} {80}},\ \bibinfo
  {pages} {045106} (\bibinfo {year} {2009})}\BibitemShut {NoStop}%
\bibitem [{\citenamefont {Ngo~Dinh}\ \emph {et~al.}(2010)\citenamefont
  {Ngo~Dinh}, \citenamefont {Bagrets},\ and\ \citenamefont
  {Mirlin}}]{Dinh.2010}%
  \BibitemOpen
  \bibfield  {author} {\bibinfo {author} {\bibfnamefont {S.}~\bibnamefont
  {Ngo~Dinh}}, \bibinfo {author} {\bibfnamefont {D.~A.}\ \bibnamefont
  {Bagrets}}, \ and\ \bibinfo {author} {\bibfnamefont {A.~D.}\ \bibnamefont
  {Mirlin}},\ }\href {\doibase 10.1103/PhysRevB.81.081306} {\bibfield
  {journal} {\bibinfo  {journal} {Phys. Rev. B}\ }\textbf {\bibinfo {volume}
  {81}},\ \bibinfo {pages} {081306(R)} (\bibinfo {year} {2010})}\BibitemShut
  {NoStop}%
\bibitem [{\citenamefont {Prosen}\ and\ \citenamefont
  {Pi{\v{z}}orn}(2008)}]{Prosen2008}%
  \BibitemOpen
  \bibfield  {author} {\bibinfo {author} {\bibfnamefont {T.}~\bibnamefont
  {Prosen}}\ and\ \bibinfo {author} {\bibfnamefont {I.}~\bibnamefont
  {Pi{\v{z}}orn}},\ }\href {\doibase 10.1103/PhysRevLett.101.105701} {\bibfield
   {journal} {\bibinfo  {journal} {Phys. Rev. Lett.}\ }\textbf {\bibinfo
  {volume} {101}},\ \bibinfo {pages} {105701} (\bibinfo {year}
  {2008})}\BibitemShut {NoStop}%
\bibitem [{\citenamefont {Prosen}\ and\ \citenamefont
  {{{\v{Z}}unkovi{\v{c}}}}(2010)}]{Prosen2010a}%
  \BibitemOpen
  \bibfield  {author} {\bibinfo {author} {\bibfnamefont {T.}~\bibnamefont
  {Prosen}}\ and\ \bibinfo {author} {\bibfnamefont {B.}~\bibnamefont
  {{{\v{Z}}unkovi{\v{c}}}}},\ }\href {\doibase 10.1088/1367-2630/12/2/025016}
  {\bibfield  {journal} {\bibinfo  {journal} {New J. Phys.}\ }\textbf {\bibinfo
  {volume} {12}},\ \bibinfo {pages} {025016} (\bibinfo {year}
  {2010})}\BibitemShut {NoStop}%
\bibitem [{\citenamefont {Prosen}(2011)}]{Prosen2011a}%
  \BibitemOpen
  \bibfield  {author} {\bibinfo {author} {\bibfnamefont {T.}~\bibnamefont
  {Prosen}},\ }\href {\doibase 10.1103/PhysRevLett.107.137201} {\bibfield
  {journal} {\bibinfo  {journal} {Phys. Rev. Lett.}\ }\textbf {\bibinfo
  {volume} {107}},\ \bibinfo {pages} {137201} (\bibinfo {year}
  {2011})}\BibitemShut {NoStop}%
\bibitem [{\citenamefont {Prosen}(2014)}]{Prosen2014}%
  \BibitemOpen
  \bibfield  {author} {\bibinfo {author} {\bibfnamefont {T.}~\bibnamefont
  {Prosen}},\ }\href {\doibase 10.1103/PhysRevLett.112.030603} {\bibfield
  {journal} {\bibinfo  {journal} {Phys. Rev. Lett.}\ }\textbf {\bibinfo
  {volume} {112}},\ \bibinfo {pages} {030603} (\bibinfo {year}
  {2014})}\BibitemShut {NoStop}%
\bibitem [{\citenamefont {Ribeiro}\ and\ \citenamefont
  {Vieira}(2015{\natexlab{a}})}]{Ribeiro2014e}%
  \BibitemOpen
  \bibfield  {author} {\bibinfo {author} {\bibfnamefont {P.}~\bibnamefont
  {Ribeiro}}\ and\ \bibinfo {author} {\bibfnamefont {V.~R.}\ \bibnamefont
  {Vieira}},\ }\href {\doibase 10.1103/PhysRevB.92.100302} {\bibfield
  {journal} {\bibinfo  {journal} {Phys. Rev. B}\ }\textbf {\bibinfo {volume}
  {92}},\ \bibinfo {pages} {100302(R)} (\bibinfo {year}
  {2015}{\natexlab{a}})},\ \Eprint {http://arxiv.org/abs/1412.8486}
  {arXiv:1412.8486} \BibitemShut {NoStop}%
\bibitem [{\citenamefont {Ribeiro}\ and\ \citenamefont
  {Vieira}(2015{\natexlab{b}})}]{Ribeiro2015f}%
  \BibitemOpen
  \bibfield  {author} {\bibinfo {author} {\bibfnamefont {P.}~\bibnamefont
  {Ribeiro}}\ and\ \bibinfo {author} {\bibfnamefont {V.~R.}\ \bibnamefont
  {Vieira}},\ }in\ \href {\doibase 10.1142/9789814740371_0005} {\emph {\bibinfo
  {booktitle} {Symmetry, Spin Dynamics and the Properties of Nanostructures}}}\
  (\bibinfo  {publisher} {World Scientific},\ \bibinfo {year} {2015})\ pp.\
  \bibinfo {pages} {86--111}\BibitemShut {NoStop}%
\bibitem [{\citenamefont {Bertini}\ \emph {et~al.}(2016)\citenamefont
  {Bertini}, \citenamefont {Collura}, \citenamefont {De~Nardis},\ and\
  \citenamefont {Fagotti}}]{PhysRevLett.117.207201}%
  \BibitemOpen
  \bibfield  {author} {\bibinfo {author} {\bibfnamefont {B.}~\bibnamefont
  {Bertini}}, \bibinfo {author} {\bibfnamefont {M.}~\bibnamefont {Collura}},
  \bibinfo {author} {\bibfnamefont {J.}~\bibnamefont {De~Nardis}}, \ and\
  \bibinfo {author} {\bibfnamefont {M.}~\bibnamefont {Fagotti}},\ }\href
  {\doibase 10.1103/PhysRevLett.117.207201} {\bibfield  {journal} {\bibinfo
  {journal} {Phys. Rev. Lett.}\ }\textbf {\bibinfo {volume} {117}},\ \bibinfo
  {pages} {207201} (\bibinfo {year} {2016})}\BibitemShut {NoStop}%
\bibitem [{\citenamefont {Castro-Alvaredo}\ \emph {et~al.}(2016)\citenamefont
  {Castro-Alvaredo}, \citenamefont {Doyon},\ and\ \citenamefont
  {Yoshimura}}]{PhysRevX.6.041065}%
  \BibitemOpen
  \bibfield  {author} {\bibinfo {author} {\bibfnamefont {O.~A.}\ \bibnamefont
  {Castro-Alvaredo}}, \bibinfo {author} {\bibfnamefont {B.}~\bibnamefont
  {Doyon}}, \ and\ \bibinfo {author} {\bibfnamefont {T.}~\bibnamefont
  {Yoshimura}},\ }\href {\doibase 10.1103/PhysRevX.6.041065} {\bibfield
  {journal} {\bibinfo  {journal} {Phys. Rev. X}\ }\textbf {\bibinfo {volume}
  {6}},\ \bibinfo {pages} {041065} (\bibinfo {year} {2016})}\BibitemShut
  {NoStop}%
\bibitem [{\citenamefont {Arrigoni}\ \emph {et~al.}(2013)\citenamefont
  {Arrigoni}, \citenamefont {Knap},\ and\ \citenamefont {von~der
  Linden}}]{Arrigoni2013}%
  \BibitemOpen
  \bibfield  {author} {\bibinfo {author} {\bibfnamefont {E.}~\bibnamefont
  {Arrigoni}}, \bibinfo {author} {\bibfnamefont {M.}~\bibnamefont {Knap}}, \
  and\ \bibinfo {author} {\bibfnamefont {W.}~\bibnamefont {von~der Linden}},\
  }\href {\doibase 10.1103/PhysRevLett.110.086403} {\bibfield  {journal}
  {\bibinfo  {journal} {Phys. Rev. Lett.}\ }\textbf {\bibinfo {volume} {110}},\
  \bibinfo {pages} {086403} (\bibinfo {year} {2013})}\BibitemShut {NoStop}%
\bibitem [{\citenamefont {Puel}\ \emph {et~al.}(2019)\citenamefont {Puel},
  \citenamefont {Chesi}, \citenamefont {Kirchner},\ and\ \citenamefont
  {Ribeiro}}]{Puel-Chesi-Kirchner-Ribeiro-2019}%
  \BibitemOpen
  \bibfield  {author} {\bibinfo {author} {\bibfnamefont {T.~O.}\ \bibnamefont
  {Puel}}, \bibinfo {author} {\bibfnamefont {S.}~\bibnamefont {Chesi}},
  \bibinfo {author} {\bibfnamefont {S.}~\bibnamefont {Kirchner}}, \ and\
  \bibinfo {author} {\bibfnamefont {P.}~\bibnamefont {Ribeiro}},\ }\href
  {\doibase 10.1103/PhysRevLett.122.235701} {\bibfield  {journal} {\bibinfo
  {journal} {Phys. Rev. Lett.}\ }\textbf {\bibinfo {volume} {122}},\ \bibinfo
  {pages} {235701} (\bibinfo {year} {2019})}\BibitemShut {NoStop}%
\bibitem [{\citenamefont {Thouless}(1969)}]{Thouless.69}%
  \BibitemOpen
  \bibfield  {author} {\bibinfo {author} {\bibfnamefont {D.~J.}\ \bibnamefont
  {Thouless}},\ }\href {\doibase 10.1103/PhysRev.187.732} {\bibfield  {journal}
  {\bibinfo  {journal} {Phys. Rev.}\ }\textbf {\bibinfo {volume} {187}},\
  \bibinfo {pages} {732} (\bibinfo {year} {1969})}\BibitemShut {NoStop}%
\bibitem [{\citenamefont {Bar}\ and\ \citenamefont
  {Mukamel}(2014)}]{PhysRevLett.112.015701}%
  \BibitemOpen
  \bibfield  {author} {\bibinfo {author} {\bibfnamefont {A.}~\bibnamefont
  {Bar}}\ and\ \bibinfo {author} {\bibfnamefont {D.}~\bibnamefont {Mukamel}},\
  }\href {\doibase 10.1103/PhysRevLett.112.015701} {\bibfield  {journal}
  {\bibinfo  {journal} {Phys. Rev. Lett.}\ }\textbf {\bibinfo {volume} {112}},\
  \bibinfo {pages} {015701} (\bibinfo {year} {2014})}\BibitemShut {NoStop}%
\bibitem [{\citenamefont {Fronczak}\ \emph {et~al.}(2016)\citenamefont
  {Fronczak}, \citenamefont {Fronczak},\ and\ \citenamefont
  {Krawiecki}}]{PhysRevE.93.012124}%
  \BibitemOpen
  \bibfield  {author} {\bibinfo {author} {\bibfnamefont {A.}~\bibnamefont
  {Fronczak}}, \bibinfo {author} {\bibfnamefont {P.}~\bibnamefont {Fronczak}},
  \ and\ \bibinfo {author} {\bibfnamefont {A.}~\bibnamefont {Krawiecki}},\
  }\href {\doibase 10.1103/PhysRevE.93.012124} {\bibfield  {journal} {\bibinfo
  {journal} {Phys. Rev. E}\ }\textbf {\bibinfo {volume} {93}},\ \bibinfo
  {pages} {012124} (\bibinfo {year} {2016})}\BibitemShut {NoStop}%
\bibitem [{\citenamefont {Angl\`es~d'Auriac}\ and\ \citenamefont
  {Igl\'oi}(2016)}]{PhysRevE.94.062126}%
  \BibitemOpen
  \bibfield  {author} {\bibinfo {author} {\bibfnamefont {J.-C.}\ \bibnamefont
  {Angl\`es~d'Auriac}}\ and\ \bibinfo {author} {\bibfnamefont {F.}~\bibnamefont
  {Igl\'oi}},\ }\href {\doibase 10.1103/PhysRevE.94.062126} {\bibfield
  {journal} {\bibinfo  {journal} {Phys. Rev. E}\ }\textbf {\bibinfo {volume}
  {94}},\ \bibinfo {pages} {062126} (\bibinfo {year} {2016})}\BibitemShut
  {NoStop}%
\bibitem [{\citenamefont {Juh\'asz}\ and\ \citenamefont
  {Igl\'oi}(2017)}]{PhysRevE.95.022109}%
  \BibitemOpen
  \bibfield  {author} {\bibinfo {author} {\bibfnamefont {R.}~\bibnamefont
  {Juh\'asz}}\ and\ \bibinfo {author} {\bibfnamefont {F.}~\bibnamefont
  {Igl\'oi}},\ }\href {\doibase 10.1103/PhysRevE.95.022109} {\bibfield
  {journal} {\bibinfo  {journal} {Phys. Rev. E}\ }\textbf {\bibinfo {volume}
  {95}},\ \bibinfo {pages} {022109} (\bibinfo {year} {2017})}\BibitemShut
  {NoStop}%
\bibitem [{\citenamefont {Alert}\ \emph {et~al.}(2017)\citenamefont {Alert},
  \citenamefont {Tierno},\ and\ \citenamefont {Casademunt}}]{Alert12906}%
  \BibitemOpen
  \bibfield  {author} {\bibinfo {author} {\bibfnamefont {R.}~\bibnamefont
  {Alert}}, \bibinfo {author} {\bibfnamefont {P.}~\bibnamefont {Tierno}}, \
  and\ \bibinfo {author} {\bibfnamefont {J.}~\bibnamefont {Casademunt}},\
  }\href {\doibase 10.1073/pnas.1712584114} {\bibfield  {journal} {\bibinfo
  {journal} {Proceedings of the National Academy of Sciences}\ }\textbf
  {\bibinfo {volume} {114}},\ \bibinfo {pages} {12906} (\bibinfo {year}
  {2017})}\BibitemShut {NoStop}%
\bibitem [{\citenamefont {{\v{Z}}unkovi{\v{c}}}\ and\ \citenamefont
  {Prosen}(2010)}]{_unkovi__2010}%
  \BibitemOpen
  \bibfield  {author} {\bibinfo {author} {\bibfnamefont {B.}~\bibnamefont
  {{\v{Z}}unkovi{\v{c}}}}\ and\ \bibinfo {author} {\bibfnamefont
  {T.}~\bibnamefont {Prosen}},\ }\href {\doibase
  10.1088/1742-5468/2010/08/p08016} {\bibfield  {journal} {\bibinfo  {journal}
  {Journal of Statistical Mechanics: Theory and Experiment}\ }\textbf {\bibinfo
  {volume} {2010}},\ \bibinfo {pages} {P08016} (\bibinfo {year}
  {2010})}\BibitemShut {NoStop}%
\bibitem [{\citenamefont {Ajisaka}\ \emph {et~al.}(2014)\citenamefont
  {Ajisaka}, \citenamefont {Barra},\ and\ \citenamefont
  {{\v{Z}}unkovi{\v{c}}}}]{Ajisaka_2014}%
  \BibitemOpen
  \bibfield  {author} {\bibinfo {author} {\bibfnamefont {S.}~\bibnamefont
  {Ajisaka}}, \bibinfo {author} {\bibfnamefont {F.}~\bibnamefont {Barra}}, \
  and\ \bibinfo {author} {\bibfnamefont {B.}~\bibnamefont
  {{\v{Z}}unkovi{\v{c}}}},\ }\href {\doibase 10.1088/1367-2630/16/3/033028}
  {\bibfield  {journal} {\bibinfo  {journal} {New Journal of Physics}\ }\textbf
  {\bibinfo {volume} {16}},\ \bibinfo {pages} {033028} (\bibinfo {year}
  {2014})}\BibitemShut {NoStop}%
\bibitem [{\citenamefont {Lieb}\ \emph
  {et~al.}(1961{\natexlab{a}})\citenamefont {Lieb}, \citenamefont {Schultz},\
  and\ \citenamefont {Mattis}}]{Lieb1961}%
  \BibitemOpen
  \bibfield  {author} {\bibinfo {author} {\bibfnamefont {E.}~\bibnamefont
  {Lieb}}, \bibinfo {author} {\bibfnamefont {T.}~\bibnamefont {Schultz}}, \
  and\ \bibinfo {author} {\bibfnamefont {D.}~\bibnamefont {Mattis}},\ }\href
  {\doibase 10.1016/0003-4916(61)90115-4} {\bibfield  {journal} {\bibinfo
  {journal} {Annals of Physics}\ }\textbf {\bibinfo {volume} {16}},\ \bibinfo
  {pages} {407} (\bibinfo {year} {1961}{\natexlab{a}})}\BibitemShut {NoStop}%
\bibitem [{\citenamefont {Kitaev}(2001)}]{Kitaev2001}%
  \BibitemOpen
  \bibfield  {author} {\bibinfo {author} {\bibfnamefont {A.~Y.}\ \bibnamefont
  {Kitaev}},\ }\href
  {http://iopscience.iop.org/article/10.1070/1063-7869/44/10S/S29/meta}
  {\bibfield  {journal} {\bibinfo  {journal} {Phys. Usp.}\ }\textbf {\bibinfo
  {volume} {44}},\ \bibinfo {pages} {131} (\bibinfo {year} {2001})}\BibitemShut
  {NoStop}%
\bibitem [{\citenamefont {Meier}\ and\ \citenamefont
  {Loss}(2003)}]{PhysRevLett.90.167204}%
  \BibitemOpen
  \bibfield  {author} {\bibinfo {author} {\bibfnamefont {F.}~\bibnamefont
  {Meier}}\ and\ \bibinfo {author} {\bibfnamefont {D.}~\bibnamefont {Loss}},\
  }\href {\doibase 10.1103/PhysRevLett.90.167204} {\bibfield  {journal}
  {\bibinfo  {journal} {Phys. Rev. Lett.}\ }\textbf {\bibinfo {volume} {90}},\
  \bibinfo {pages} {167204} (\bibinfo {year} {2003})}\BibitemShut {NoStop}%
\bibitem [{\citenamefont {Nakata}\ \emph {et~al.}(2017)\citenamefont {Nakata},
  \citenamefont {Simon},\ and\ \citenamefont {Loss}}]{Nakata_2017}%
  \BibitemOpen
  \bibfield  {author} {\bibinfo {author} {\bibfnamefont {K.}~\bibnamefont
  {Nakata}}, \bibinfo {author} {\bibfnamefont {P.}~\bibnamefont {Simon}}, \
  and\ \bibinfo {author} {\bibfnamefont {D.}~\bibnamefont {Loss}},\ }\href
  {\doibase 10.1088/1361-6463/aa5b09} {\bibfield  {journal} {\bibinfo
  {journal} {Journal of Physics D: Applied Physics}\ }\textbf {\bibinfo
  {volume} {50}},\ \bibinfo {pages} {114004} (\bibinfo {year}
  {2017})}\BibitemShut {NoStop}%
\bibitem [{\citenamefont {Hoffman}\ \emph {et~al.}(2018)\citenamefont
  {Hoffman}, \citenamefont {Loss},\ and\ \citenamefont
  {Tserkovnyak}}]{hoffman2018superfluid}%
  \BibitemOpen
  \bibfield  {author} {\bibinfo {author} {\bibfnamefont {S.}~\bibnamefont
  {Hoffman}}, \bibinfo {author} {\bibfnamefont {D.}~\bibnamefont {Loss}}, \
  and\ \bibinfo {author} {\bibfnamefont {Y.}~\bibnamefont {Tserkovnyak}},\
  }\href@noop {} {\enquote {\bibinfo {title} {Superfluid transport in quantum
  spin chains},}\ } (\bibinfo {year} {2018}),\ \Eprint
  {http://arxiv.org/abs/1810.11470} {arXiv:1810.11470 [cond-mat.mes-hall]}
  \BibitemShut {NoStop}%
\bibitem [{\citenamefont {Shen}\ \emph {et~al.}(2020)\citenamefont {Shen},
  \citenamefont {Hoffman},\ and\ \citenamefont {Trif}}]{shen2020theory}%
  \BibitemOpen
  \bibfield  {author} {\bibinfo {author} {\bibfnamefont {P.-X.}\ \bibnamefont
  {Shen}}, \bibinfo {author} {\bibfnamefont {S.}~\bibnamefont {Hoffman}}, \
  and\ \bibinfo {author} {\bibfnamefont {M.}~\bibnamefont {Trif}},\ }\href@noop
  {} {\enquote {\bibinfo {title} {Theory of topological spin josephson
  junction},}\ } (\bibinfo {year} {2020}),\ \Eprint
  {http://arxiv.org/abs/1912.11458} {arXiv:1912.11458 [cond-mat.mes-hall]}
  \BibitemShut {NoStop}%
\bibitem [{\citenamefont {Barouch}\ and\ \citenamefont
  {McCoy}(1971)}]{Barouch.71}%
  \BibitemOpen
  \bibfield  {author} {\bibinfo {author} {\bibfnamefont {E.}~\bibnamefont
  {Barouch}}\ and\ \bibinfo {author} {\bibfnamefont {B.~M.}\ \bibnamefont
  {McCoy}},\ }\href {\doibase 10.1103/PhysRevA.3.786} {\bibfield  {journal}
  {\bibinfo  {journal} {Phys. Rev. A}\ }\textbf {\bibinfo {volume} {3}},\
  \bibinfo {pages} {786} (\bibinfo {year} {1971})}\BibitemShut {NoStop}%
\bibitem [{\citenamefont {Lieb}\ \emph
  {et~al.}(1961{\natexlab{b}})\citenamefont {Lieb}, \citenamefont {Schultz},\
  and\ \citenamefont {Mattis}}]{lieb1961two}%
  \BibitemOpen
  \bibfield  {author} {\bibinfo {author} {\bibfnamefont {E.}~\bibnamefont
  {Lieb}}, \bibinfo {author} {\bibfnamefont {T.}~\bibnamefont {Schultz}}, \
  and\ \bibinfo {author} {\bibfnamefont {D.}~\bibnamefont {Mattis}},\
  }\href@noop {} {\bibfield  {journal} {\bibinfo  {journal} {Annals of
  Physics}\ }\textbf {\bibinfo {volume} {16}},\ \bibinfo {pages} {407}
  (\bibinfo {year} {1961}{\natexlab{b}})}\BibitemShut {NoStop}%
\bibitem [{\citenamefont {Sachdev}(1996)}]{sachdev1996universal}%
  \BibitemOpen
  \bibfield  {author} {\bibinfo {author} {\bibfnamefont {S.}~\bibnamefont
  {Sachdev}},\ }\href@noop {} {\bibfield  {journal} {\bibinfo  {journal}
  {Nuclear Physics B}\ }\textbf {\bibinfo {volume} {464}},\ \bibinfo {pages}
  {576} (\bibinfo {year} {1996})}\BibitemShut {NoStop}%
\bibitem [{\citenamefont {Banchi}\ \emph {et~al.}(2014)\citenamefont {Banchi},
  \citenamefont {Giorda},\ and\ \citenamefont {Zanardi}}]{Banchi2013}%
  \BibitemOpen
  \bibfield  {author} {\bibinfo {author} {\bibfnamefont {L.}~\bibnamefont
  {Banchi}}, \bibinfo {author} {\bibfnamefont {P.}~\bibnamefont {Giorda}}, \
  and\ \bibinfo {author} {\bibfnamefont {P.}~\bibnamefont {Zanardi}},\ }\href
  {\doibase 10.1103/PhysRevE.89.022102} {\bibfield  {journal} {\bibinfo
  {journal} {Phys. Rev. E}\ }\textbf {\bibinfo {volume} {89}},\ \bibinfo
  {pages} {022102} (\bibinfo {year} {2014})},\ \Eprint
  {http://arxiv.org/abs/1305.4527} {arXiv:1305.4527} \BibitemShut {NoStop}%
\bibitem [{\citenamefont {Medvedyeva}\ and\ \citenamefont
  {Kehrein}(2014)}]{Medvedyeva2014}%
  \BibitemOpen
  \bibfield  {author} {\bibinfo {author} {\bibfnamefont {M.~V.}\ \bibnamefont
  {Medvedyeva}}\ and\ \bibinfo {author} {\bibfnamefont {S.}~\bibnamefont
  {Kehrein}},\ }\href {\doibase 10.1103/PhysRevB.90.205410} {\bibfield
  {journal} {\bibinfo  {journal} {Phys. Rev. B}\ }\textbf {\bibinfo {volume}
  {90}},\ \bibinfo {pages} {1} (\bibinfo {year} {2014})},\ \Eprint
  {http://arxiv.org/abs/1406.1408} {arXiv:1406.1408} \BibitemShut {NoStop}%
\bibitem [{\citenamefont {Klich}\ and\ \citenamefont
  {Levitov}(2009)}]{PhysRevLett.102.100502}%
  \BibitemOpen
  \bibfield  {author} {\bibinfo {author} {\bibfnamefont {I.}~\bibnamefont
  {Klich}}\ and\ \bibinfo {author} {\bibfnamefont {L.}~\bibnamefont
  {Levitov}},\ }\href {\doibase 10.1103/PhysRevLett.102.100502} {\bibfield
  {journal} {\bibinfo  {journal} {Phys. Rev. Lett.}\ }\textbf {\bibinfo
  {volume} {102}},\ \bibinfo {pages} {100502} (\bibinfo {year}
  {2009})}\BibitemShut {NoStop}%
\bibitem [{\citenamefont {Song}\ \emph {et~al.}(2010)\citenamefont {Song},
  \citenamefont {Rachel},\ and\ \citenamefont {Le~Hur}}]{PhysRevB.82.012405}%
  \BibitemOpen
  \bibfield  {author} {\bibinfo {author} {\bibfnamefont {H.~F.}\ \bibnamefont
  {Song}}, \bibinfo {author} {\bibfnamefont {S.}~\bibnamefont {Rachel}}, \ and\
  \bibinfo {author} {\bibfnamefont {K.}~\bibnamefont {Le~Hur}},\ }\href
  {\doibase 10.1103/PhysRevB.82.012405} {\bibfield  {journal} {\bibinfo
  {journal} {Phys. Rev. B}\ }\textbf {\bibinfo {volume} {82}},\ \bibinfo
  {pages} {012405} (\bibinfo {year} {2010})}\BibitemShut {NoStop}%
\bibitem [{\citenamefont {Song}\ \emph {et~al.}(2011)\citenamefont {Song},
  \citenamefont {Flindt}, \citenamefont {Rachel}, \citenamefont {Klich},\ and\
  \citenamefont {Le~Hur}}]{PhysRevB.83.161408}%
  \BibitemOpen
  \bibfield  {author} {\bibinfo {author} {\bibfnamefont {H.~F.}\ \bibnamefont
  {Song}}, \bibinfo {author} {\bibfnamefont {C.}~\bibnamefont {Flindt}},
  \bibinfo {author} {\bibfnamefont {S.}~\bibnamefont {Rachel}}, \bibinfo
  {author} {\bibfnamefont {I.}~\bibnamefont {Klich}}, \ and\ \bibinfo {author}
  {\bibfnamefont {K.}~\bibnamefont {Le~Hur}},\ }\href {\doibase
  10.1103/PhysRevB.83.161408} {\bibfield  {journal} {\bibinfo  {journal} {Phys.
  Rev. B}\ }\textbf {\bibinfo {volume} {83}},\ \bibinfo {pages} {161408}
  (\bibinfo {year} {2011})}\BibitemShut {NoStop}%
\bibitem [{\citenamefont {Song}\ \emph {et~al.}(2012)\citenamefont {Song},
  \citenamefont {Rachel}, \citenamefont {Flindt}, \citenamefont {Klich},
  \citenamefont {Laflorencie},\ and\ \citenamefont
  {Le~Hur}}]{PhysRevB.85.035409}%
  \BibitemOpen
  \bibfield  {author} {\bibinfo {author} {\bibfnamefont {H.~F.}\ \bibnamefont
  {Song}}, \bibinfo {author} {\bibfnamefont {S.}~\bibnamefont {Rachel}},
  \bibinfo {author} {\bibfnamefont {C.}~\bibnamefont {Flindt}}, \bibinfo
  {author} {\bibfnamefont {I.}~\bibnamefont {Klich}}, \bibinfo {author}
  {\bibfnamefont {N.}~\bibnamefont {Laflorencie}}, \ and\ \bibinfo {author}
  {\bibfnamefont {K.}~\bibnamefont {Le~Hur}},\ }\href {\doibase
  10.1103/PhysRevB.85.035409} {\bibfield  {journal} {\bibinfo  {journal} {Phys.
  Rev. B}\ }\textbf {\bibinfo {volume} {85}},\ \bibinfo {pages} {035409}
  (\bibinfo {year} {2012})}\BibitemShut {NoStop}%
\bibitem [{\citenamefont {Sachdev}(2011)}]{Sachdev-2011}%
  \BibitemOpen
  \bibfield  {author} {\bibinfo {author} {\bibfnamefont {S.}~\bibnamefont
  {Sachdev}},\ }\href@noop {} {\emph {\bibinfo {title} {Quantum Phase
  Transitions}}},\ \bibinfo {edition} {2nd}\ ed.\ (\bibinfo  {publisher}
  {Cambridge University Press},\ \bibinfo {year} {2011})\BibitemShut {NoStop}%
\bibitem [{\citenamefont {Castin}(2007)}]{Castin2007}%
  \BibitemOpen
  \bibfield  {author} {\bibinfo {author} {\bibfnamefont {Y.}~\bibnamefont
  {Castin}},\ }in\ \href
  {https://hal.archives-ouvertes.fr/hal-00122049/document} {\emph {\bibinfo
  {booktitle} {Proceedings of the International School of Physics ``Enrico
  Fermi''}}},\ \bibinfo {editor} {edited by\ \bibinfo {editor} {\bibfnamefont
  {M.}~\bibnamefont {Inguscio}}, \bibinfo {editor} {\bibfnamefont
  {W.}~\bibnamefont {Ketterle}}, \ and\ \bibinfo {editor} {\bibfnamefont
  {C.}~\bibnamefont {Salomon}}}\ (\bibinfo  {publisher} {SIF, Bologna, Italy},\
  \bibinfo {year} {2007})\ pp.\ \bibinfo {pages} {289--349}\BibitemShut
  {NoStop}%
\bibitem [{\citenamefont {Vidal}\ \emph {et~al.}(2003)\citenamefont {Vidal},
  \citenamefont {Latorre}, \citenamefont {Rico},\ and\ \citenamefont
  {Kitaev}}]{Vidal.03}%
  \BibitemOpen
  \bibfield  {author} {\bibinfo {author} {\bibfnamefont {G.}~\bibnamefont
  {Vidal}}, \bibinfo {author} {\bibfnamefont {J.~I.}\ \bibnamefont {Latorre}},
  \bibinfo {author} {\bibfnamefont {E.}~\bibnamefont {Rico}}, \ and\ \bibinfo
  {author} {\bibfnamefont {A.}~\bibnamefont {Kitaev}},\ }\href {\doibase
  10.1103/PhysRevLett.90.227902} {\bibfield  {journal} {\bibinfo  {journal}
  {Phys. Rev. Lett.}\ }\textbf {\bibinfo {volume} {90}},\ \bibinfo {pages}
  {227902} (\bibinfo {year} {2003})}\BibitemShut {NoStop}%
\bibitem [{\citenamefont {Its}\ and\ \citenamefont {Korepin}(2009)}]{Its.2009}%
  \BibitemOpen
  \bibfield  {author} {\bibinfo {author} {\bibfnamefont {A.~R.}\ \bibnamefont
  {Its}}\ and\ \bibinfo {author} {\bibfnamefont {V.~E.}\ \bibnamefont
  {Korepin}},\ }\href {\doibase 10.1007/s10955-009-9835-9} {\bibfield
  {journal} {\bibinfo  {journal} {Journal of Statistical Physics}\ }\textbf
  {\bibinfo {volume} {137}},\ \bibinfo {pages} {1014} (\bibinfo {year}
  {2009})}\BibitemShut {NoStop}%
\bibitem [{\citenamefont {Calabrese}\ and\ \citenamefont
  {Cardy}(2004)}]{Calabrese.04}%
  \BibitemOpen
  \bibfield  {author} {\bibinfo {author} {\bibfnamefont {P.}~\bibnamefont
  {Calabrese}}\ and\ \bibinfo {author} {\bibfnamefont {J.}~\bibnamefont
  {Cardy}},\ }\href {http://stacks.iop.org/1742-5468/2004/i=06/a=P06002}
  {\bibfield  {journal} {\bibinfo  {journal} {J. Stat. Mech.}\ }\textbf
  {\bibinfo {volume} {2004}},\ \bibinfo {pages} {P06002} (\bibinfo {year}
  {2004})}\BibitemShut {NoStop}%
\bibitem [{\citenamefont {Eisler}\ and\ \citenamefont
  {Zimbor\'as}(2014)}]{Eisler.14}%
  \BibitemOpen
  \bibfield  {author} {\bibinfo {author} {\bibfnamefont {V.}~\bibnamefont
  {Eisler}}\ and\ \bibinfo {author} {\bibfnamefont {Z.}~\bibnamefont
  {Zimbor\'as}},\ }\href {\doibase 10.1103/PhysRevA.89.032321} {\bibfield
  {journal} {\bibinfo  {journal} {Phys. Rev. A}\ }\textbf {\bibinfo {volume}
  {89}},\ \bibinfo {pages} {032321} (\bibinfo {year} {2014})}\BibitemShut
  {NoStop}%
\bibitem [{\citenamefont {Ribeiro}(2017)}]{Ribeiro.17}%
  \BibitemOpen
  \bibfield  {author} {\bibinfo {author} {\bibfnamefont {P.}~\bibnamefont
  {Ribeiro}},\ }\href {\doibase 10.1103/PhysRevB.96.054302} {\bibfield
  {journal} {\bibinfo  {journal} {Phys. Rev. B}\ }\textbf {\bibinfo {volume}
  {96}},\ \bibinfo {pages} {054302} (\bibinfo {year} {2017})}\BibitemShut
  {NoStop}%
\end{thebibliography}%

\end{document}